\documentclass[preprint, showpacs, preprintnumbers, amsmath, amssymb, superscriptaddress, nofootinbib]{revtex4}
\usepackage{graphicx,color}
\usepackage{amsmath,amssymb,slashed}
\usepackage{url}
\usepackage{epstopdf}

\newcommand{\be}{\begin{equation}}
\newcommand{\ee}{\end{equation}}
\newcommand{\bea}{\begin{eqnarray}}
\newcommand{\eea}{\end{eqnarray}}

\newcommand{\crn}{\nonumber \\}
\newcommand{\non}{\nonumber}

\newcommand{\fr}{\frac}

\newcommand{\bc}{\begin{center}}
\newcommand{\ec}{\end{center}}

\newcommand {\ba}{\begin{array}}
\newcommand {\ea}{\end{array}}
\newcommand{\ben}{\begin{enumerate}}
\newcommand{\een}{\end{enumerate}}

\newcommand{\crb}[1]{{\color{blue}#1}}

\usepackage{bm}
\usepackage{dcolumn}

\begin{document}

\title{ One-loop contributions to decays $e_b\to e_a \gamma$  and $(g-2)_{e_a}$ anomalies,  and Ward identity}

\author{L.T. Hue}
\email{lethohue@vlu.edu.vn}

\author{H. N. Long}
\email{hoangngoclong@vlu.edu.vn}
\affiliation{Subatomic Physics Research Group, Science and Technology Advanced Institute, Van Lang University, Ho Chi Minh City, Vietnam}
\affiliation{Faculty of Applied Technology, School of Technology, Van Lang University, Ho Chi Minh City, Vietnam}
\author{V. H. Binh} 
\affiliation{Institute of Physics, Vietnam Academy of Science and Technology,	10 Dao Tan, Ba Dinh, Hanoi, Vietnam}
\author{H.~L.~T.~Mai}
\affiliation{Faculty of Physics Science, Can Tho Medical College, Nguyen Van Cu Street, Can Tho, Vietnam}
\author{T. Phong Nguyen \footnote{corresponding author}}
\email{thanhphong@ctu.edu.vn}
\affiliation{Department of Physics, Can Tho University, 	3/2 Street, Can Tho, Vietnam}
\begin{abstract}
In this paper, we will present  analytic formulas to express  one-loop contributions to lepton flavor violating decays $e_b\to e_a \gamma$, which are also relevant to the  anomalous dipole magnetic moments  of charged leptons $e_a$. These formulas were computed in the unitary gauge,  using the  well-known Passarino-Veltman notations.  We also show that our results are  consistent with those  calculated previously in the   't Hooft-Veltman gauge, or in the limit of zero lepton masses.  At the one-loop level, we show that the appearance of  fermion-scalar-vector type diagrams in the unitary gauge  will violate the Ward Identity relating to an external photon. As a result, the validation of the Ward Identity  guarantees that  the photon always couples with two identical particles in  an arbitrary triple coupling vertex containing a photon.  
\end{abstract} 
\maketitle
 \section{\label{intro} Introduction}
 \allowdisplaybreaks
 The lepton sector is one of the most interesting objects  for experiments to search for  new physics (NP) beyond the  prediction of the standard model (SM).  For example, the evidence of neutrino oscillation confirms that the  SM must be extended. Recently, the  experimental data of  anomalous magnetic moments (AMM) of charged leptons $(g-2)_{e_a}/2\equiv a_{e_a}$ has  been updated, where the deviation between SM prediction  and the lasted experiment data  for  muon is   \cite{Muong-2:2021ojo}   
\begin{equation} 
\label{eq_damu}
\Delta a^{\mathrm{NP}}_{\mu}\equiv  a^{\mathrm{exp}}_{\mu} -a^{\mathrm{SM}}_{\mu} =\left(251 \pm 59\right)  \times 10^{-11},
\end{equation}
 corresponding to  the $4.2\sigma$ deviation  from  standard model (SM) prediction \cite{Aoyama:2020ynm} combined from various contributions \cite{Davier:2010nc, Davier:2017zfy, Keshavarzi:2018mgv, Colangelo:2018mtw, Hoferichter:2019mqg, Davier:2019can, Keshavarzi:2019abf, Kurz:2014wya, Melnikov:2003xd, Masjuan:2017tvw, Colangelo:2017fiz, Hoferichter:2018kwz, Gerardin:2019vio, Bijnens:2019ghy, Colangelo:2019uex, Colangelo:2014qya, Blum:2019ugy, Aoyama:2012wk, Aoyama:2019ryr, Czarnecki:2002nt, Gnendiger:2013pva}.
 For the electron anomaly, the deviation between SM and experiment is  $1.6\sigma$ discrepancy   \cite{Morel:2020dww}.

On the other hand,  $\Delta a_{e,\mu}$ are strongly constrained by the  experimental data obtained from searching for the charged lepton flavor violating  (cLFV) decays  $e_b\rightarrow e_a\gamma$ are~\cite{MEG:2016leq, BaBar:2009hkt}:
\begin{align}
	\label{eq_ebagaex}
	\mathrm{Br}(\tau\rightarrow \mu\gamma)&<4.4\times 10^{-8}, \; 
	\mathrm{Br}(\tau\rightarrow e\gamma) <3.3\times 10^{-8}, \;
	\mathrm{Br}(\mu\rightarrow e\gamma) < 4.2\times 10^{-13}.
\end{align}
This important property was discussed previously, for example see  discussions for a general estimation  in Ref.  \cite{Crivellin:2018qmi}, and many  particular models  beyond the standard model (BSM)  \cite{Lindner:2016bgg, Dorsner:2020aaz, Hue:2021xap, Hue:2021xzl, Hong:2022xjg, Li:2022zap}.  General formulas  expressing simultaneously both one-loop contributions to AMM and cLFV amplitudes were introduced in the limits of  new heavy scalar and/or gauge boson exchanges $m_B^2 \gg m^2_{a}$ with $m_a$ being the mass of a charged lepton  $e_a=e,\mu,\tau$ \cite{Crivellin:2018qmi}. Other  calculations in the unitary gauge  were discussed  \cite{Yu:2021suw, Leveille:1977rc} for  the one-loop contributions to $a_{e_a}$ with $m_{a}\neq0$, without the relations with the cLFV amplitudes.   The analytic one-loop formulas for  cLFV amplitudes  calculated in  the 't Hooft Feynman (HF) gauge   were  also shown in Ref. \cite{Lavoura:2003xp}, using the notations of the Passarino-Veltman (PV) functions \cite{Passarino:1978jh, tHooft:1978jhc} with $m_{a}\neq m_{b}$. The approximate formulas with $m_{a}=m_{b}=0$ were introduced and consistent with those given in Ref. \cite{Crivellin:2018qmi}, as shown particularly  in Ref. \cite{Hue:2017lak} for  3-3-1 models. The general analytic formulas of these PV functions were introduced for numerical investigations. They are consistent with the results generated by  LoopTools \cite{Hahn:1998yk}, which can be transformed  into   other PV notations  implemented in the Fortran numerical package \textit{Collier} \cite{Denner:2016kdg}, used to investigate  cLFV  decays in a two Higgs doublet model (2HDM)  \cite{Jurciukonis:2021izn}.  Many particular expressions to compute the AMM and/or cLFV decay amplitudes predicted by   different particular BSM  were constructed  \cite{Lindner:2016bgg}. The relations  among them can be checked by using suitable transformations, starting from the set of particular PV notations in this work.  On the other hand, in a discussion on analytic formulas for one-loop contributions  to AMM,   a class of fermion-scalar-vector ($FSV$)  diagrams  consisting of a  photon coupling  with two different physical particles, namely one scalar and one gauge boson,  were considered even in the unitary gauge \cite{Yu:2021suw}. It leads us to whether the Ward identity (WI) for the external photon is still valid with the presence of this diagram type.   We emphasize that the general results for one-loop contributions to decays $e_b\to e_a \gamma$ and AMM of leptons introduced in  many previous works do not include these $FSV$ diagrams.  Moreover, they imply the existence of the triple photon  coupling  with  two distinguishable  physical particles that has  never been mentioned previously.
 In particular,   many works introducing general one-loop contributions for  AMM of charged leptons \cite{Leveille:1977rc, Lindner:2016bgg, Crivellin:2018qmi},  or decays relating with photon such as cLFV decays  $e_b\to e_a\gamma$ \cite{Lavoura:2003xp, Lindner:2016bgg, Crivellin:2018qmi}, loop-induced Higgs decays  $h\to \gamma \gamma$ \cite{Gunion:1989we, Bunk:2013uea}, $h\to Z\gamma, f\bar{f} \gamma$ \cite{Bunk:2013uea, Hue:2017cph, Phan:2021xwc, VanOn:2021myp}, quark decays $q\to q'\gamma$, $\dots$. Excluding the $FSV$ vertex type will reduce a huge number of  related one- and two-loop diagrams as well as  confirm the validation of general one-loop calculation introduced previously.   

In this work, we will show precisely  the important steps to derive the one-loop contributions to both AMM and cLFV decays. The calculation is performed by hand, which is consistent with another cross-checking using FORM package \cite{Vermaseren:2000nd}. The final formulas are expressed exactly in terms of the PV functions defined by LoopTools.   The results are then easy to change into all the other available forms  using suitable transformations. The conventions of the PV-functions are very convenient to derive the exact formulas before solving particular pure mathematical problems.  We also determine  contributions arising from a new form of photon coupling with vector bosons such as  leptoquarks  and confirm the consistency  between our results and those introduced in Ref. \cite{Bunk:2013uea, Barbieri:2015yvd, Biggio:2016wyy}.

Our paper is organized as follows.  Section \ref{intro} explains our aim of this work. Section \ref{sec_formulas} introduces notations and important formulas to establish the relations between AMM and cLFV amplitudes. Section \ref{sec_discuss} shows  discussions to confirm the consistency  of our results and previous works,  and the validation of the WI for the relevant analytic formulas. Section \ref{sec_conclusion} summarizes main features of our work. Finally, we provide  many appendices  showing precisely many intermediate steps  and notations to derive the final results mentioned in this work, including the analytic formulas of the PV functions consistent with LoopTools given in  appendix~\ref{app_PVLT}.

\section{ \label{sec_formulas} General amplitudes and notations}
\begin{figure}[ht]
	\centering
	\includegraphics[width=14cm]{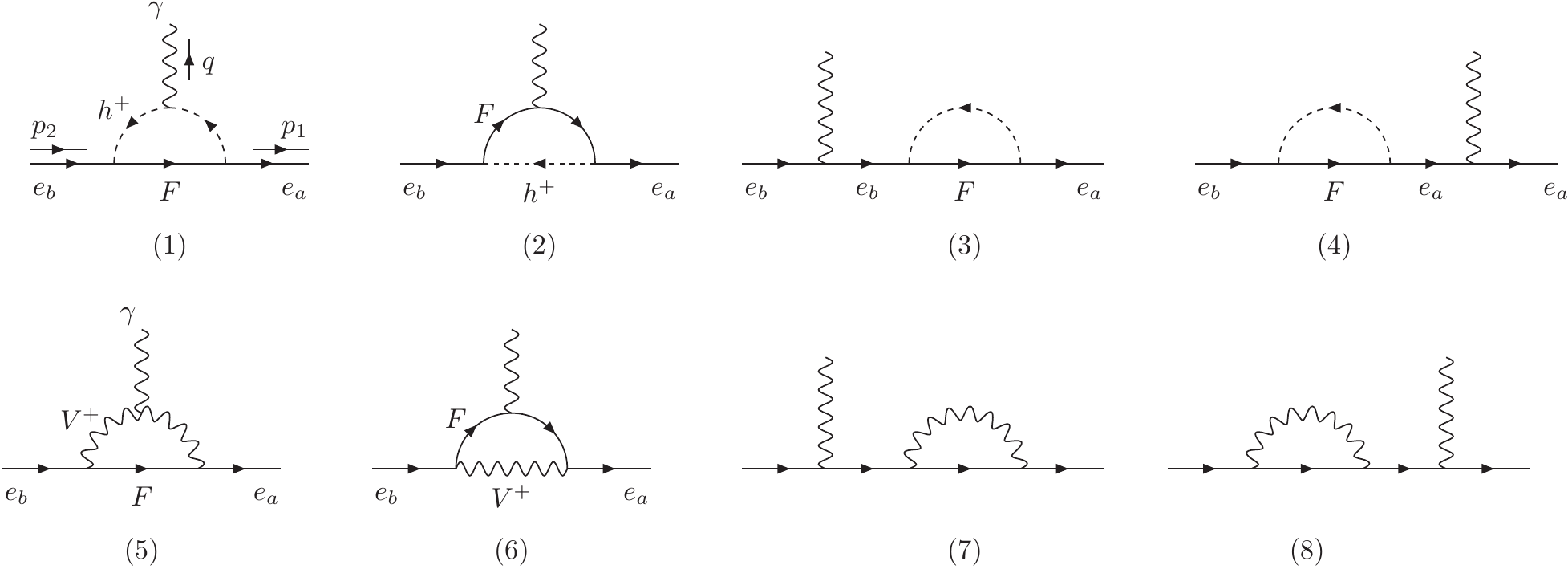}
	\caption{Feynman diagrams for one-loop contribution to $a_{e_a}$ and cLFV amplitudes $e_b\to e_a \gamma$  in the unitary gauge.  
		\label{fig_eab}}
\end{figure}

It is well-known that analytic formulas of one-loop contributions to the cLFV amplitudes  $e_b(p_2)\rightarrow e_a(p_1)\gamma(q)$  and AMM  of  SM charged leptons $e_a$  can be presented in the same expressions, see for example Ref. \cite{Crivellin:2018qmi} corresponding to the presence of new heavy particles in BSM.  Possible one-loop Feynman diagrams contributing to $a_{e_a}$ and cLFV decay amplitudes $e_b\to e_a \gamma$ in  BSM are shown in Fig.~\ref{fig_eab}, where $F$ is a fermion coupling with the SM charged lepton $e_a=e,\mu,\tau$; and the boson  $B=h, V$ is a scalar or gauge boson, respectively. 
For a detailed calculation, precise conventions for external momenta and propagators are presented in appendix \ref{app_detailedStep}. We note here  that Ref. \cite{Yu:2021suw} argues another type of $FSV$  one-loop diagrams giving new contributions to the  AMM. They will be discussed in detail in this work. 

Firstly, we adopt the Lagrangian generating one-loop diagrams in Fig. \ref{fig_eab}, namely  \cite{Crivellin:2018qmi}
\begin{align}
\mathcal{L}_h &=\overline{F}(g_{a, Fh}^{L}P_L +g_{a,Fh}^{R}P_R) e_a h +\mathrm{h.c.},	\label{eq_LFh}
\\
\mathcal{L}_V& =\overline{F}\gamma^{\mu}  ( g_{a,FV}^{L}P_L +g_{a, FV}^{R}  P_R)e_aV_{\mu} +\mathrm{h.c.} \label{eq_LFV},
\end{align}
where the fermion $F$ and the  boson $B=V_{\mu},h$ have  electric charges $Q_F$ and $Q_B$,    and  masses $m_F$ and  $m_B$, respectively.   These Lagrangians \eqref{eq_LFh} and \eqref{eq_LFV} are consistent with those in Ref. \cite{Lavoura:2003xp}. Moreover, the photon couplings  with all physical particles should be mentioned clearly, as given in Ref. \cite{Lavoura:2003xp}, i.e., we will adopt the Feynman rules that the  photon always couples with two identical physical particles, as  given in table \ref{t_AXX},
\begin{table}[h]
	\begin{tabular}{|c|c|c|c|c|c|}
		\hline
		Vertex & Coupling & 	Vertex  & Couplings &Vertex  & Couplings\\
		\hline
		$A^{\mu}(p_0)V^{\nu}(p_+)V^{*\lambda}(p_-)$&$-ieQ_V\Gamma_{\mu \nu \lambda}(p_0,p_+,p_-) $&$A^{\mu}h(p_+)h^*(p_-)$&$ ieQ_h(p_+-p_-)_{\mu}$ & $A^{\mu}\overline{F}F$& $ieQ_F\gamma_{\mu}$\\
		\hline
	\end{tabular}
	\caption{Feynman rules for cubic couplings of photon $A^{\mu}$, where $p_{0,\pm}$ are incoming momenta into the relevant vertex.   
		\label{t_AXX}}
\end{table} 
where $\Gamma_{\mu \nu \lambda}(p_0,p_+,p_-) = g_{\mu\nu} (p_0 -p_+)_{\lambda} +g_{\nu \lambda} (p_+ -p_-)_{\mu} +g_{ \lambda \mu} (p_- -p_0)_{\nu}$ is the standard form.  The  more general form of $\Gamma_{\mu \nu \lambda}(p_0,p_+,p_-)$ introduced in Refs. \cite{Bunk:2013uea, Barbieri:2015yvd, Biggio:2016wyy} will be discussed in detail  
later.

All couplings listed in Lagrangians \eqref{eq_LFh}, \eqref{eq_LFV},  and table \ref{t_AXX} result in the following   form factors relevant with one-loop contributions:
\begin{align} \label{eq_Boson}
	c_{R B}^{ab}= &\dfrac{e}{16\pi^2}g_{a,FB}^{L*} g_{b,FB}^{R} m_F  \times \dfrac{f_B(x_B) +Q_Fg_B(x_B)}{m_B^2} 
	\nonumber \\
	&+ \dfrac{e}{16\pi^2}\left(m_{b} g_{a,FB}^{L*} g_{b, FB}^{L} +m_{a} g_{a,FB}^{R*} g_{b, FB}^{R}\right) \times \dfrac{\tilde{f}_B (x_B) +Q_F \tilde{g}_B(x_B)}{m_B^2},
\end{align}
where $x_B \equiv m_F^2/m_B^2$. The four scalar functions $f_B(x)$, $g_B(x)$,  $\tilde{f}_B(x)$, and $\tilde{g}_B(x)$ are listed in Eq. \eqref{eq_fgx} of  appendix \ref{app_PVLT}, as the approximate formulas in the limit $m_a,m_b\ll m_B$. The formula in Eq.  \eqref{eq_Boson} does not contain contributions  from the $FSV$ diagrams mentioned in Ref. \cite{Yu:2021suw}, because of the absence of  photon coupling $AVh$. The corresponding formulas of AMM and cLFV decay rates are:
\begin{align}
	a_{e_a} &\equiv  -\dfrac{2m_{a}}{e}\left(c^{aa}_R + c^{aa*}_R\right) = -\dfrac{4m_{a}}{e}\mathrm{Re}[ c^{aa}_R], \label{eq_aea1}	
\\ \mathrm{Br} (e_b\to e_a \gamma)&= \frac{m^3_b}{4\pi \Gamma_b}\left( \left|c^{ab}_R\right|^2 + \left|c^{ba}_R\right|^2\right),
	\end{align}
where $m_a$, $m_b$, and $\Gamma_b$ are the masses and total decay width of the leptons $e_a$, $e_b$, and 
\begin{align}
	\label{eq_cabR}	
c^{ab}_R &\equiv \sum_{B,F} c^{ab}_{RB}. 
  \end{align}

The amplitude  for a vertex $\bar{e}_a e_aA_{\mu}$ in Ref. \cite{Peskin:1995ev}  is consistent with the following  form presenting both AMM and cLFV amplitudes \cite{Escribano:1996wp,Eidelman:2016aih} 
\begin{equation}\label{eq_eeAeff}
	i\mathcal{M}=-ie \overline{u_a}(p_1)\left[ \gamma^{\mu}F_1 -\frac{\sigma^{\mu\nu}q_{\nu}}{2m_{a}} \left(iF_2 + \gamma^5 F_3\right)\right]u_b(p_2)\varepsilon^*_{\mu},
\end{equation}
where $\sigma^{\mu\nu}\equiv \frac{i}{2}\left[ \gamma^{\mu} \gamma^{\nu} - \gamma^{\nu} \gamma^{\mu}\right]$;  $F_{1,2,3}$ are scalar form factors; $\varepsilon^*_{\mu}$ and $q_{\nu}$ is the polarized vector of the external photon. The derivation of Eq. \eqref{eq_eeAeff} respecting the WI from the most general form was explained clearly in Ref. \cite{Eidelman:2016aih}. The form factors $F_{2,3}$ get contributions only from loop corrections. They relate with the well-known experimental quantities called the anomalous magnetic moment $a_{e_a}$ and electric dipole moment $d_{e_a}$ for $b=a$, respectively. Specifically, we have $	F_{1}=1$ for the on-shell photon, and
\begin{equation}\label{eq_ga}
\quad a_{e_a}=F_2; \quad  d_{e_a}=-\frac{e}{2m_{a}}F_3. 
\end{equation}
Regarding the LFV decay $ e_b\rightarrow e_a \gamma$ the amplitude can also be written in the same form \cite{Cheng:1984vwu, Lavoura:2003xp}, suggesting that $F_2$ can be calculated based on the  one-loop corrections to LFV decays. In particular, the second term of the amplitude \eqref{eq_eeAeff} can be expanded as follows \cite{Hue:2017lak} 
\begin{align}
	\mathcal{M}&=(2p_1.\varepsilon^*)\overline{u_a} \left( C_{(ab)L} P_L +C_{(ab)R} P_R\right)u_b 
	 +\overline{u_a} \left[D_{(ab)L} \slashed{\varepsilon}^* P_L +D_{(ab)R} \slashed{\varepsilon}^* P_R\right]u_b,
\end{align}
where $m_{a}=m_{b}$ and we can prove that $C_{(ab)L}P_L +C_{(ab)R} P_R =\frac{e}{2m_{a}}(F_2 -i\gamma^5 F_3)$.  The WI for the external photon gives
\begin{equation}\label{eq_DLR}
D_{(ab)L}= -(m_{b}C_{(ab)R}  +m_{a}C_{(ab)L}), \; D_{(ab)R} = -(m_{b} C_{(ab)L}  +m_{a}C_{(ab)R}).
\end{equation}  
We note that although WI does not require the condition of on-shell photon $q^2=0$ in general, it was also used to derive the two relations given in  Eq. \eqref{eq_DLR}, which simplify our calculation in the unitary gauge \footnote{We thank the referee for reminding us this point}. The general case of $q^2=0$ is beyond our scope, see Ref. \cite{Jurciukonis:2021izn} for a detailed discussion of this case in the 2HDM framework. 
The hermiticity that   $C_{(aa)R}=C^*_{(aa)L}$ \cite{Eidelman:2016aih} gives
\begin{align}
	\label{eq_FCLR}
	a_{e_a}&= \frac{m_{a}(C_{(aa)L} + C_{(aa)R})}{e} =\frac{2m_{a} \mathrm{Re}{[C_{(aa)L,R}]} }{e},\crn 
	d_{e_a}&=i(C_{(aa)R} -C_{(aa)L})=\mathrm{Im}{[C_{(aa)L}]}= -\mathrm{Im}{[C_{(aa)R}]}.
\end{align}
Hence, the following relations between two different notations must be satisfied:
\begin{equation}\label{eq_relationCbaR}
	c^{ab}_{R}= -\frac{1}{2} C_{(ab)R}  \; \mathrm{and} \; c^{ba}_{R}= -\frac{1}{2} C_{(ab)L}.
\end{equation}
 From the above discussion, we see that one-loop contributions to the $a_{e_a}$ and $d_{e_a}$ can be written in terms of well-known PV functions,  see detailed discussions in Ref.  \cite{Hue:2017lak} or general formula introduced for calculations of the cLFV decay rates  \cite{Lavoura:2003xp}, with the identification that $\sigma_{L,R}\equiv -C_{(ab)L,R}$.  In the limit of $0\simeq m_{a},m_{b}\ll m_B$, the numerical values of  $a_{e_a}$ can be evaluated using the numerical packages such as  LoopTools \cite{Hahn:1998yk} or Collier \cite{Denner:2016kdg}. Although the exact analytic formulas of one-loop three-point functions  presented in Ref.~\cite{Hue:2017lak} can not be applied to calculate $a_{e_a}$, the limit of $m_b\to m_a$ can be used to solve this problem. The analytic formulas of $a_{e_a}$  were introduced completely in Ref. \cite{Yu:2021suw}.    
 
 Because of the relations in Eq. \eqref{eq_DLR}, only $C_{(ab)L,R}$ is  needed to determine $a_{e_a}$ and Br$(e_b\to e_a \gamma)$. Because all two-point diagrams  give contributions to just   $D_{(ab)L,R}$,  $C_{(ab)L,R}$ are calculated by considering only three-point diagrams. In this work, the analytic formulas of  $D_{(ab)L,R}$ will be determined directly from all diagrams in Fig. \ref{fig_eab} to  check the validation of the WI in the presence of the $FSV$.  

The analytic formulas  for one-loop  contributions to   the cLFV decay amplitudes  presented 
in this work are more general than  the results introduced in Ref.~\cite{Hue:2017lak}  for  general 3-3-1 models.  Many important steps in our calculations were shown  in  appendix~\ref{app_detailedStep}. Using this unitary gauge, the  assumption for a particular form of  the Goldstone boson couplings  given in Ref. \cite{Lavoura:2003xp} is unnecessary. In contrast, we use the same  photon  couplings to  other physical particles in an arbitrary BSM,  as given in table \ref{t_AXX}. Namely,   a tree-level photon coupling always contains two identical physical particles.  This implies that the contributions from the $FSV$ diagrams are not included. 

  Using the notations of PV-functions defined in appendix \ref{app_PVLT}, the $Fhh$ contributions from diagram (1)  in Fig. \ref{fig_eab}  are:
\begin{align}\label{eq_C_Lfhh}
	C^{Fhh}_{(ab)L} =& \frac{-eQ_h}{16\pi^2} \left[ m_{a}g_{a, Fh}^{L*} g_{b, Fh}^{L}  X_1^{f} 
	+  m_{b}g_{a, Fh}^{R*} g_{b,Fh}^{R} X_2^{f}  -m_F g_{a,Fh}^{R*} g_{b,Fh}^L X_0^{f} \right],
	\crn C^{Fhh}_{(ab)R} =& \frac{-eQ_h}{16\pi^2}\left[m_{a}g_{a,Fh}^{R*} g_{Fh}^{bR}X_1^{f} 
	 + m_{b}g_{a, Fh}^{L*} g_{b,Fh}^{L}X_2^{f}  -m_F g_{a, Fh}^{L*}g_{b, Fh}^{R}X_0^{f}\right],
	 %
\end{align}
where  $X^f_{0},X^f_1,\dots$ are linear combinations of the PV-functions $C_{0, 00, i, ij}(m_F^2, m_h^2,m_h^2)$ defined precisely  in appendix \ref{app_PVLT}. 

 The   diagram (2)  in Fig. \ref{fig_eab}  gives $hFF$ contributions as follows:
 \begin{align}
 	\label{eq_CLhff}
 	C^{hFF}_{(ab)L} =& \frac{-eQ_F}{16\pi^2}\left[ m_{a}g_{a,Fh}^{L*} g_{b,Fh}^{bL} X_1^{h} 
 	+   m_{b}g_{a,Fh}^{R*} g_{b,Fh}^{R}X_2^{h} +m_F g_{a,Fh}^{R*} g_{b,Fh}^L X_3^{h} \right], 
 	\crn C^{hFF}_{(ab)R} =& \frac{-eQ_F}{16\pi^2} \left[m_{a}g_{a,Fh}^{R*} g_{b,Fh}^{R}X_1^{h}  
 	+ m_{b}g_{a,Fh}^{L*} g_{b, Fh}^{L} X_2^{h} +m_F g_{a, Fh}^{L^*}g_{b,Fh}^{R}X_3^{h}\right], 
 	%
 \end{align}	
where   $X^h_{1,2,3}$ are linear combinations of  $C_{0,i,ij}(m_h^2, m_F^2, m_F^2)$.  The above results are completely consistent with those introduced in Ref. \cite{Lavoura:2003xp}, except an overall sign and the signs before the PV-functions $\bar{c}_{1,2}$,  arising   from the different definitions of the external momenta $p_i$ in the denominators of the one-loop integrals.  We also give the analytic formulas of $D^{Fhh}_{(ab)L,R}$ and $D^{hFF}_{(ab)L,R}$, used to confirm the  WI given in Eq. \eqref{eq_DLR}  for the only-scalar contributions.  The PV-functions derived from diagram (2) defined as $X^h_{i}$ are different from $X^f_{i}$  defined for  three diagrams (1), (3), and (4). In contrast, the  equal functions are denoted  as follows: 
$$ B^{(i)}_0 \equiv B^{(i)f}_0=B^{(i)h}_0= B_0(p_i^2,m_h^2,m_F^2), \;X_0\equiv X_0^f=X_0^h, \; i=1,2. $$
 The form factors $D_{(ab)L,R}$ originated from scalar contributions are: 
\begin{align}
\label{eq_Hformfactor}
 D_{(ab)L}^{Fhh}=& \frac{-eQ_H}{16\pi^2} \left\{g^{L*}_{a,Fh}g^{L}_{b,Fh} \times 2 C^f_{00}  \right\}  
\crn &
+ \frac{-eQ_e}{16\pi^2(m_a^2 -m_b^2)} \left\{ \left(m_b g^{L*}_{a,Fh}g^{R}_{b,Fh} + m_a g^{R*}_{a,Fh} g^{L}_{b,Fh} \right)m_F  \left( B^{(1)}_0 -B^{(2)}_0 \right) 
\right. \crn &- \left.  g^{L*}_{a,Fh} g^{L}_{b,Fh}\left( m^2_a B^{(1)f}_1- m^2_b B^{(2)f}_1 \right) -    m_a m_b g^{R*}_{a,Fh} g^{R}_{b,Fh} \left( B^{(1)f}_1- B^{(2)f}_1\right) \right\},
\crn D_{(ab)R}^{Fhh} =&  D_{(ab)L}^{FHH} \left[ g^L_{a,Fh} \leftrightarrow g^R_{a,Fh}, g^L_{b,Fh} \leftrightarrow g^R_{b,Fh} \right],
%
\crn D_{(ab)L}^{hFF} =& -\frac{eQ_F}{16 \pi^2} \left\{ g^{L*}_{a,Fh} g^{L}_{b,Fh} \left[m_F^2 C^h_0 +(2-d) C^h_{00} -m_a^2X_1^h - m_b^2 X_2^h \right]
\right.\crn&\left. + g^{R*}_{a,Fh} g^{R}_{b,Fh} m_a m_bX_0 +    \left[ g^{R*}_{a,Fh} g^{L}_{b,Fh} m_a + g^{L*}_{a,Fh} g^{R}_{b,Fh} m_b\right] m_F C^h_0 \right\},
\crn D_{(ab)R}^{hFF} =& D_{(ab)L}^{hFF}\left[ g^L_{a,Fh} \leftrightarrow g^R_{a,Fh}, g^L_{b,Fh} \leftrightarrow g^R_{b,Fh} \right],
\end{align}
where   $X^h_{1,2,3}$ are linear combinations of  $C^h_{0,i,ij} \equiv C_{0,i,ij}(m_h^2, m_F^2, m_F^2)$, $C^f_{00} \equiv C_{00} ( m_F^2,m_h^2, m_h^2)$, and $B^{(i)f}_1 \equiv B^{(i)}_1(m_F^2,m_h^2)$ given in Eq. \eqref{eq_Bix}.

It is noted that the $Fhh$ contributions are the sum of three diagrams (1), (3), and (4), while the $hFF$ contributions are  from only diagram (2). We emphasize that the electric charge conservation $Q_F=Q_h+Q_e$ is one of the necessary requirements  to guarantee the WI given in Eq. \eqref{eq_DLR}, see a detailed proof in appendix \ref{app_detailedStep}. We can see this crudely from the necessary condition that   div$[D^{hFF}_{(ab)L}]+ \mathrm{div}[D^{Fhh}_{(ab)L}]\sim g_{a}^{L*}g^L_b (Q_e+Q_h-Q_F)=0$ and div$[D^{hFF}_{(ab)R}]+ \mathrm{div}[D^{Fhh}_{(ab)R}] \sim g_{a}^{R*}g^R_b(Q_e+Q_h-Q_F)=0$. This conclusion supports completely the only case of electric conservation among the remaining ones  mentioned in  Ref. \cite{Lavoura:2003xp}. 

Regarding  Lagrangian \eqref{eq_LFV}, which results in four diagrams in the second line of Fig.  \ref{fig_eab},   diagram (5) gives the following $FVV$ contributions:
\begin{align}
	\label{eq_CLFVV}	
	C_{(ab)L}^{FVV} 
	= -\frac{eQ_V}{16\pi^2} &
	\left\{ g_{a,FV}^{R*}g_{b,FV}^{L} m_F \left[ 3X_3^f +\frac{1}{2m_V^2} \right]   
	- g_{a,FV}^{L*}g_{b, FV}^{R} m_F\times \frac{  m_{a}m_{b}}{ m_V^2} X^f_{012}   
	\right. \crn & + g_{a, FV}^{L*}g_{b,FV}^{L}m_{a} \left[2 (X^f_1 -X^f_3)   +  \frac{m_F^2 X^f_{01} +m_{b}^2X^f_2}{m_V^2}  \right]
	\crn & \left. + g_{a, FV}^{R*}g_{b, FV}^{R}m_{b} \left[ 2 (X^f_2  -X^f_3)  +  \frac{m_F^2 X^f_{02} +m_{a}^2X^f_1}{m_V^2} \right] \right\},
\end{align}
where  $X^{f}_{i}$ is the linear combinations of $C_{0,ij}(m_F^2, m_V^2, m_V^2)$, given in Eq. \eqref{eq_Xix},  
 and 
\begin{align}
	\label{eq_CRFVV}	
	C_{(ab)R}^{FVV} 
	= -\frac{eQ_V}{16\pi^2} &
	\left\{ g_{a, FV}^{L*}g_{b, FV}^{R} m_F \left[ 3X^f_3 +\frac{1}{2m_V^2}  \right]  - g_{a, FV}^{R*}g_{b, FV}^{L} m_F\times \frac{  m_{a}m_{b}}{ m_V^2} X^f_{012}  
	\right. \crn & +g_{a, FV}^{R*}g_{b, FV}^{R}m_{a} \left[ 2 (X^f_1 -X^f_3)  +\frac{m_F^2 X^f_{01}+ m_{b}^2 X^f_2}{m_V^2}  \right]
	\crn & \left. +g_{a, FV}^{L*}g_{b,FV}^{L}m_{b} \left[2 (X^f_2 -X^f_3)      +\frac{m_F^2X^f_{02}+ m_{a}^2 X^f_1}{m_V^2}  \right] \right\}.
\end{align} 
Diagram (6) gives $VFF$ contributions:
	\begin{align}
	\label{eq_VFFCL}	
	C_{(ab)L}^{VFF}= -\frac{eQ_F}{16\pi^2} &\left\{  m_{a} g_{a, FV}^{L*} g_{b, FV}^{L}  \left[ \frac{}{} 2 X^v_{01} 
	+\frac{m_F^2\left( X^v_1  -X^v_3\right)+ m_{b}^2X^v_2 }{m_V^2} \right]
	\right. \crn&\; + m_{b}g_{a, FV}^{R*} g_{b,FV}^{R} \left[ \frac{}{}2 X^v_{02} 
	%
	+\frac{m_F^2\left( X^v_2  -X^v_3\right)+ m_{a}^2 X^v_1}{m_V^2} \right]
	\crn&\; - g_{a,FV}^{R*} g_{b,FV}^{L} m_F \left[  4 X_0   
	%
	+ \frac{m_{a}^2X^v_1  + m_{b}^2X^v_2-  m^2_{F}X^v_3}{m_V^2} \right]
	\crn&\left. \;-g_{a, FV}^{L*}g_{b, FV}^{R}  \frac{m_{a} m_{b}}{m_V^2}\times m_F (X^v_{12} -X^v_3)   \right\},
\end{align}
where  all $X_{i}^v$ are expressed in terms of PV functions $C^{VFF}_{0,i,ij}=C_{0,i,ij}(m_V^2, m_F^2, m_F^2)$,  and 
\begin{align}
	\label{eq_VFFCR}	
	C_{(ab)R}^{VFF}= -\frac{eQ_F}{16\pi^2} &\left\{  m_{a} g_{a, FV}^{R*} g_{b,FV}^{R} \left[ 2 X^v_{01} 
	+\frac{m_F^2\left( X^v_1  -X^v_3\right)+ m_{b}^2X^v_2}{m_V^2} \right]
	\right. \crn&\; + m_{b}g_{a, FV}^{L*} g_{b, FV}^{L} \left[ 2 X^v_{02}
	%
+\frac{m_F^2\left( X^v_2 -X^v_3\right)+ m_{a}^2X^v_1}{m_V^2} \right]
	\crn&\; - g_{a, FV}^{L*} g_{b, FV}^{R} m_F \left[  4 X^v_0   
	%
 + \frac{m_{a}^2X^v_1  + m_{b}^2X^v_2 - m^2_{F}X^v_3}{m_V^2} \right]
	\crn&\left. \;-g_{a,FV}^{R*}g_{b, FV}^{L}  \frac{m_{b} m_{a}}{m_V^2}\times m_F (X^v_{12} -X^v_3)   \right\}.  
	%
\end{align}
Finally, using  the simple notations $g^{L,R}_a\equiv g^{L,R}_{a,FV}$,  the formulas of  $D_{(ab)L}$ and $D_{(ab)R}$ are 
\begin{align}
D^{(78)}_{(ab)L}=&	D^{(7)}_{(ab)L}+ D^{(8)}_{(ab)L}
\crn = &  \frac{e Q_e}{ 16\pi^2 (m_a^2 -m_b^2)} \left\{ \frac{}{} \left(  g^{L*}_{a} g^{R}_{b} m_b + g^{R*}_{a} g^{L}_{b} m_a\right) 3m_F \left[ B^{(1)}_0 -B^{(2)}_0\right]
\right. \crn &- \left. m_b \left( m_a g^{R*}_{a} g^{R}_{b} + m_b g^{L*}_{a} g^{L}_{b}\right)   \left[ \left(2+  \frac{m_F^2 +m_b^2}{m_V^2}\right) B^{(2)v}_1  +\frac{A_0(m_V^2) +2m_F^2B^{(1)}_0}{m_V^2} +1 \right] 
\right. \crn &+\left. m_a \left( m_b g^{R*}_{a} g^{R}_{b} + m_a g^{L*}_{a} g^{L}_{b}\right)   \left[ \left(2+  \frac{m_F^2 +m_a^2}{m_V^2}\right) B^{(1) v}_1 +\frac{A_0(m_V^2) +2m_F^2B^{(2)}_0}{m_V^2} +1 \right]\right\}, \label{eq_DabLR78}
\\ D^{(78)}_{(ab)R}=& D^{(78)}_{(ab)L} \left[ g^L_{a} \leftrightarrow g^R_{a}, \; g^L_{b} \leftrightarrow g^R_{b}\right].
%
\crn D_{(ab)L}^{FVV}=& -\frac{eQ_V}{16\pi^2} \left\{\frac{}{}   g_{a}^{L*} g_{b}^{L}  \left[\frac{}{} 2(d-2)C^f_{00} +2(m_a^2+m_b^2)X^f_3  
\right.\right.\crn&\left. \left. -\frac{1}{m_V^2} \left(  m_F^2(B^{(1)}_0 +B^{(2)}_0 -2C^f_{00} ) +A_0(m_V^2) +m_a^2 B^{(1) f}_1+m_b^2 B^{(2) f}_1  \right)\right]
\right. \crn&\; + g_{a}^{R*} g_{b}^{R} m_a m_{b} \left[4X^f_3 +\frac{2 C^f_{00}}{m_V^2} \right]
%
 + g_{a}^{R*} g_{b}^{L} \times m_a m_F \left[ 3C^f_0 -\frac{1}{2m_V^2} + \frac{m_b^2 X^f_{012}}{m_V^2}\right]
\crn&\left. \; +g_{a}^{L*}g_{b}^{R} \times m_b m_F \left[ 3C^f_0 -\frac{1}{2m_V^2} + \frac{m_a^2 X^f_{012}}{m_V^2}\right] \right\},
\crn D_{(ab)R}^{FVV}=& C_{(ab)L}^{FVV} \left[ g^{L}_{a} \leftrightarrow g^{R}_{a}, g^{L}_{b} \leftrightarrow g^{R}_{a}\right] \label{eq_FVVDLhand},
\end{align}
where  all $X_{i}^f$ are expressed in terms of PV functions $C^f_{0,ij} \equiv C_{0,ij}(m_F^2, m_V^2,m_V^2)$ and $B^{(i)f}_1$ is given in Eq. \eqref{eq_Bix}.

 The remaining formulas of $D_{(ab)L,R}$ from diagram (6) of Fig. \ref{fig_eab} are  
\begin{align}
\label{eq_DLR6}
D^{VFF}_{(ab)L}=&  \frac{eQ_F}{16\pi^2} \left\{ g^{L*}_{a} g^{L}_{b} \left[\frac{}{} -2m_F^2 C_{0} +(d-2)^2 C_{00}^v +2m_a^2 X^v_{01}+2m_b^2 X^v_{02} 
\right.\right.\crn&\left.\left. \qquad \qquad  \qquad \quad  -\frac{1}{m_V^2} \left[(2-d)m_F^2 C^v_{00} +A_0(m_V^2)  +m_F^2 \left( B^{(1)}_0 +B^{(2)}_0 \right)
\right.\right. \right. \crn&\left.\left. \left.  \qquad \qquad  \qquad \qquad \qquad - m_a^2\left(B^{(1) v}_0 +B^{(1)}_1 \right) -m_b^2 \left( B^{(2) v}_0 +B^{(2)v}_1\right) +m_a^2m_b^2 X_0
\right.\right. \right. \crn&\left.\left. \left. \qquad \qquad  \qquad \qquad \qquad 
-m_F^2 \left( (m_a^2 +m_b^2-m_F^2)C_0 +m_a^2 X^v_1 +m_b^2X^v_2\right) \frac{}{} \right] \right] 
\right.\crn&\left. \qquad \quad + g^{R*}_ag^{R}_b m_am_b \left[ 2X_0  -\frac{1}{m_V^2}\left((2-d) C^v_{00} +m_F^2 X^v_3 -m_a^2 X^v_1 -m_b^2 X^v_2 \right)\right]
\right.\crn &\left.\qquad \quad  + \frac{g^{R*}_ag^{L}_b  m_am_F}{m_V^2} \left[ -B^{(1)v}_1 +(2-d)C_{00} -m_a^2 X^v_1 +m_b^2(X^v_3 -X^v_2) \right] 
\right.\crn &\left.\qquad \quad  + \frac{g^{L*}_ag^{R}_b  m_bm_F}{m_V^2} \left[ -B^{(2) v}_1 +(2-d)C^v_{00} -m_b^2 X^v_2 +m_a^2 (X^v_3 -X^v_1) \right]  \right\},
\crn D^{VFF}_{(ab)R}= & D^{VFF}_{(ab)L} \left[ g_a^L\leftrightarrow g_a^R,\; g_b^L\leftrightarrow g_b^R\right],
\end{align}
where  all $X_{i}^v$ are expressed in terms of PV functions $C^v_{0,ij} \equiv C_{0,ij}(m_V^2, m_F^2, m_F^2)$ and $B^{(i)v}_1$ is given in Eq. \eqref{eq_Bix}.

We note that all results presented here are crosschecked by FORM package \cite{Vermaseren:2000nd}, using intermediate steps given in appendix \ref{app_detailedStep}.  There is a property that $C^X_{(ab)R}=C^X_{(ab)L}\left[ g_a^L\leftrightarrow g_a^R,\; g_b^L\leftrightarrow g_b^R\right]$ for all $X=Fhh,hFF,FVV,VFF$. The above results of  one-loop contribution to $C_{(ab)L,R}$ are totally consistent with those introduced in Ref. \cite{Lavoura:2003xp}, after some 
transformations of notations presented in appendix \ref{app_special}. In the limit of $m_h^2,m_V^2\gg m^2_a,m^2_b$, i.e., $m^2_a/m_B^2,m^2_b/m_h^2\simeq 0$ with $B=h,V$, we get  consistent results with those given in Ref. \cite{Crivellin:2018qmi, Freitas:2014pua, Stockinger:2006zn}.  To derive the above results for gauge boson exchanges, we start with many important features different from those mentioned in Ref. \cite{Lavoura:2003xp}, namely: i) we do   
not use the typical form of couplings relating to Goldstone bosons going along with the presence of new gauge bosons, ii) we have to use the massless property of the  on-shell photon $q^2=0$, iii) to confirm the WI for all diagrams given in  Fig. \ref{fig_eab}, we need the charge conservation law corresponding to the Lagrangian \eqref{fig_eab}: $Q_F=Q_V +Q_e$.   Therefore, our calculation is  another independent approach to confirm the result given in Ref. \cite{Lavoura:2003xp}. The details of the calculation to confirm the WI for all one-loop contributions  are given in appendix \ref{app_detailedStep}.  We remind that our results are derived from the  photon couplings listed in the table \ref{t_AXX}, and  do not contain the contributions from the FSV diagrams.  In the following, we pay attention to the possibility of adding the FSV diagrams or  the new forms of the photon couplings.  

\section{\label{sec_discuss} Discussion on  WI and previous results}
\subsection{WI to constrain the form of  photon couplings}
Now we focus on the feature that the WI of the on-shell photon will constrain strongly the forms of the cubic  photon couplings with two physical particles in a renormalized Lagrangian. Now we consider the existence of the photon coupling types at tree level: 
\begin{align}
	\label{eq_AX12}	
	\mathcal{L}^{\gamma XX} =&  eQ_FA^{\mu}\left[ \overline{F_1}\gamma^{\mu}F_2 +\mathrm{h.c.} \right] + i eQ_hA^{\mu} \left[ \left( h_1^* \partial_{\mu}h_2 -h_2 \partial_{\mu}h_1^*\right) +\mathrm{h.c.}\right]
	\crn &   -\left[ eQ_VA^{\mu}V^{\nu}_1 V^{\lambda*}_2 \Gamma_{\mu \nu \lambda}(p_0,p_+p_-) + \mathrm{h.c.}\right] +\left[g_{\gamma hV} g_{\mu \nu}h^{-Q} A^{\mu}V^{Q\nu}+ \mathrm{h.c.}\right],
\end{align}
where all couplings are more general than those  well-known as the standard forms  given in Table \ref{t_AXX}.  In addition,  the last term   corresponds to the photon coupling to a scalar $h \equiv S$ and a gauge boson $V$  mentioned  in Eq. \eqref{eq_A-S-V}. The above Lagrangian  results in the following decays from the heavy particle to the lighter one: i) $F_2\to F_1 \gamma$, ii) $h_2\to h_1 \gamma$, iii)  $V_2\to V_1 \gamma$, and iv) $V\to h\gamma$. The WI for these decay amplitudes at tree level is $\mathcal{M}^{\mu}(X_1\to X_2 \gamma)p_{0\mu}=0$ with $p_{0\mu}$ being the external photon momentum. It can be derived that:
\begin{itemize}
	\item Using the same convention of external momenta given in Fig. \ref{fig_eab},  we have  $\mathcal{M}^{\mu}(F_2\to F_1 \gamma)q_{\mu}\sim (m_{F_2}-m_{F_1}) \overline{u}_{F_2}(p_2)u_{F_1}(p_1)=0$, where $p_0\equiv -q$. Therefore, $m_{F_2}=m_{F_1}$. This case is automatically satisfied for the tree-level AMM amplitude.
	
	\item $\mathcal{M}^{\mu}(h_2\to h_1 \gamma)p_{0\mu}\sim(p_2-p_1).(p_2+p_1)= (m^2_{h_2}-m^2_{h_1}) =0$, where all on-shell momenta are incoming the vertex $A^{\mu}h^*_1h_2$, implying  that $p_0=-(p_1+p_2)$ and $p_{1,2}^2=m^2_{h_{1,2}}$. The consequence is $m_{h_1}=m_{h_2}$. 
	\item  $\mathcal{M}^{\mu}(V \to h \gamma)p_{0\mu}\sim\varepsilon_v.p_{0} =0$, where $\varepsilon_v$ and $p_0$ are the polarization of gauge boson $V$ and the external momentum of the photon $A_{\mu}$.  Hence, the presence of a $AhV$ vertex does not automatically satisfy  the WI. One-loop contributions for all diagrams arising from this vertex  must be checked for the  validation of WI. In Ref. \cite{Yu:2021suw}, the presence of these vertices was mentioned in a Higgs triplet model (HTM). A detailed calculation in appendix E shows an opposite conclusion that this vertex vanishes at tree level \footnote{\crb{To the best of our knowledge, we have not seen any UV models beyond the SM that  have violated $U(1)_{em}$ couplings of the form $S$-$V$-$\gamma$, which is the necessary source for generating $FSV$-type diagrams.}}.
	\item  $\mathcal{M}_{\mu}(V_1 \to V_2 \gamma)p^{\mu}_0 \sim\varepsilon^{\nu}_1\varepsilon^{\lambda*}_2p^{\mu}_0 \Gamma_{\mu \nu \lambda}(p_0,p_1,p_2) =0$, where $\varepsilon_{1,2}$, and $p_{1, 2, 0}$ are the polarization of the gauge boson $V_{1,2}$ and the external momentum of the gauge bosons  $V_{1,2}$ and  photon $A_{\mu}$, respectively. We will use the following properties of the external gauge bosons $V_i (i=1,2)$ and photon:  $\varepsilon_i.p_{i}=0$, $p_0^2=0$, $p_i^2=m^2_{V_i}$, and the momentum conservation $p_0 +p_1 +p_2=0$ following notations in table \ref{t_AXX}. After some intermediate calculating steps, we have:
	\begin{align}
	\label{eq_WIV12ga}	
	\mathcal{M}_{\mu}(V_1 \to V_2 \gamma)p^{\mu}_0 \sim& (p_0.\varepsilon_1) \left[ (p_0 -p_1). \varepsilon^*_2\right] + (\varepsilon_1.\varepsilon^*_2) \left[ (p_1 -p_2).p_0 \right] +(p_0.\varepsilon^*_2) \left[ (p_2 -p_0).\varepsilon_1 \right]
	\crn =& (\varepsilon_1.\varepsilon^*_2) \left[ m^2_{V_2} -m^2_{V_1} \right]=0.
	\end{align}
Hence, $m_{V_1}= m_{V_2}$ is necessary. From this, we consider  the  more general   photon coupling  with  a gauge boson \cite{Biggio:2016wyy} describing the couplings of  a leptoquark field \cite{Barbieri:2015yvd}
\begin{align}
	\label{eq_Gamma1}
\Gamma'_{\mu\nu\lambda}(p_0,p_1,p_2) =& g_{\mu\nu} (k_v p_0 -p_1)_{\lambda} +g_{\nu \lambda} (p_1 -p_2)_{\mu} +g_{ \lambda \mu} (p_2 - k_vp_0)_{\nu}
\crn =&\Gamma_{\mu\nu\lambda}(p_0,p_1,p_2) +\delta k_v \left(g_{\mu \nu} p_{0\lambda} -g_{\lambda \mu} p_{0\nu}\right),
\end{align}
with $\delta k_v=k_v-1$ showing the deviation from the standard vertex listed in table \ref{t_AXX}. This  may change the one-loop contributions  of the diagram (5) in Fig. \ref{fig_eab}, hence  change the formulas of $C^{FVV}_{(ab)L,R}$ given in Eqs. \eqref{eq_CLFVV} and \eqref{eq_CRFVV}, respectively.  One can prove immediately that the vertex deviation 
 \begin{align}
 	\label{eq_dGamma}
\delta \Gamma_{\mu\nu\lambda}(p_0,p_1,p_2)\equiv \Gamma'_{\mu\nu\lambda}(p_0,p_1,p_2) -\Gamma_{\mu\nu\lambda}(p_0,p_1,p_2)= \delta k_v \left(g_{\mu \nu} p_{0\lambda} -g_{\lambda \mu} p_{0\nu}\right) 
 \end{align}
 guarantees the WI. The new one-loop contributions arising from    $\delta \Gamma$ are also satisfied the WI, see analytic formulas given in Eq. \eqref{eq_dCLR}. 
\end{itemize}

Now we start from the point that all results of one loop contributions  given from Eq. \eqref{eq_C_Lfhh} to Eq. \eqref{eq_DLR6} based on the standard forms of photon couplings given in table \ref{t_AXX}, where a photon always couples with two identical physical fields.  On the other hand, a recent work \cite{Yu:2021suw} assumed  the existence of a  new photon coupling kind $ASV$, which may appear in some BSM, in which the photon couples with one gauge boson $V$ and one scalar $S$.  The appearance of a boson $V$ or $S$ will  generate by itself the one-loop contributions that always guarantee the WI by the respective set of four diagrams given in Fig. \ref{fig_eab}. Hence, the two FSV diagrams must give contributions satisfying the WI themselves, namely
\begin{align}
\label{eqFSVWi}
D^{FSV}_{(ab)_L} +m_a C^{FSV}_{(ab)_L}	+m_b C^{FSV}_{(ab)_R}= D^{FSV}_{(ab)_R} +m_a C^{FSV}_{(ab)_R}	+m_b C^{FSV}_{(ab)_L}=0. 
\end{align}
As a result,  the divergent parts of $h\equiv S$ given  in appendix \ref{app_WiFSV}  for both $L$ and $R$ parts give: 
 \begin{align}
 \label{eq_divFSV}
0=& g_{\gamma hV} \left[ 2g^{L*}_{ah}g^{L}_{bV} m_F -g^{L*}_{ah} g^{R}_{bV} m_b -g^{R*}_{ah}g^{L}_{bV} m_a \right]
\crn =& g_{\gamma hV}  \left[ 2g^{R*}_{ah}g^{R}_{bV} m_F -g^{R*}_{ah}g^{L}_{bV} m_b -g^{L*}_{ah}g^{R}_{bV} m_a \right].
\end{align} 
Considering the case of $g_{\gamma hV}\neq0$. Then, all quantities $g^L_{ah}, g^R_{ah}$, $g^L_{bV}$, and $g^R_{bV}$ are zeros if at least one of them is zero.   More strictly, we require that the two  Eqs. \eqref{eqFSVWi}  must be held for both divergent and finite parts arising from  $D_{(ab)L,R}$ and $C_{(ab)L,R}$ given in appendix \ref{app_WiFSV}. Consequently,   $g_{\gamma hV}=0$, i.e.,  the  $FSV$ diagram  type does not satisfy the WI. 

Regarding the vertex deviation of the $AVV$ couplings  defined in Eq. \eqref{eq_dGamma},    the new one-loop contributions relating to $C^{FVV}_{(ab)L,R}$ and $D^{FVV}_{(ab)L,R}$ are shown in Eq. \eqref{eq_dCLR} of appendix \ref{app_detailedStep}. Our results are consistent with previous works \cite{Barbieri:2015yvd, Biggio:2016wyy}. Although they satisfy the WI,  they contain divergences. For example,  the divergent part of $\delta C^{FVV}_{L}$ is 
\begin{align}
	\label{eq_divdCLR}	
\mathrm{div}\left[ -\delta C^{FVV}_{L}\right]	 =&  \frac{\delta k_v eQ_V}{32\pi^2 m_V^2} \left[g^{L*}_a g^{L}_bm_a  +  g^{R*}_a g^{R}_b m_b -2g^{R*}_a g^{L}_b m_F  \right]. 
\end{align}
Hence, $\delta k_v=0$ is equivalent to the renormalizable  condition of  the theory, see a more detailed explanation in Ref. \cite{Biggio:2016wyy}. This confirms that the $AVV$ coupling listed in table \ref{t_AXX} is still valid for a general UV-complete model. Consequently,   $\delta C^{FVV}_{L}=0$, implying that  the results of $C^{FVV}_{(ab)L,R}$ given in Eqs. \eqref{eq_CLFVV} and \eqref{eq_CRFVV} are unchanged  for  many  renormalizable theories.

\subsection{Discussions on previous results }
It is easy to derive that $C_{(ab)L,R} = \sigma_{L,R}$ corresponding to the notations given in Ref. \cite{Lavoura:2003xp}, see a detailed explanation in appendix \ref{app_special}. This confirms a perfect consistency of the two results obtained  from different original assumptions  that we have indicated above. In addition, these  results are also consistent with those given in Ref. \cite{Crivellin:2018qmi} in the limit of heavy boson masses in the loops,  which are very useful for studying the correlations of AMM and cLFV decays. 

In some BSM, SM light quark  may play the role of the light fermions $u,d\equiv F$ in the Yukawa couplings \cite{Dorsner:2020aaz}, hence  the condition $m_F^2\gg m^2_{a},m^2_{b}$ is not held. But  numerical illustrations \cite{Hue:2017lak} to investigate cLFV decays $e_b\to e_a \gamma$  with very light neutrinos show that the case of $m^2_{F}\ll m^2_{a}$ is also valid for approximation formulas with $m^2_{a}=m^2_{b}=0$, provided $m^2_{a}, m^2_{b}\ll m^2_h,m^2_{V}$. An analytic approximation to explain this result was given in, for example, Ref. \cite{Hue:2015fbb}. 

For analytic formulas of cLFV and $a_{e_a}$ introduced in Ref. \cite{Lindner:2016bgg}, They can be changed into the form of PV-functions consistent with our results.  An  exceptional case mentioned is the coupling of a doubly charged boson with two identical leptons. For example, the Lagrangian containing couplings of  a doubly charged Higgs boson is  \cite{Lindner:2016bgg}:
\begin{align}
	\label{eq_phieab0}
	\mathcal{L}_{\mathrm{int}}= g^{ij}_{s3}\phi^{++} \overline{\ell^C_i}\ell_j +  g^{ij}_{p3}\phi^{++}  \overline{\ell^C_i} \gamma^5 \ell_j +\mathrm{h.c.},
\end{align}
where we can identify that $ g^R_{a,Fh}= g^{ij}_{s3} +g^{ij}_{p3}$ and  $g^R_{a,Fh}= g^{ij}_{s3} -g^{ij}_{p3}$. But the Feynman rules for the vertex $\overline{\ell^C_i}\ell_j \phi^{++}$ containing two identical leptons give an extra factor 2,  implying that  $C_{(ab)L,R}$ given in Eqs. \eqref{eq_C_Lfhh} and \eqref{eq_CLhff} must be added a  factor 4.   Instead of many particular formulas to calculate one-loop contributions relating to  different charged particles, the one-loop  results for $(g-2)_{e_a}$  and $e_b\to e_a \gamma$ decays can be generalized for $a_{e_a}$ with an arbitrary electric charge  $Q_F$  of a new fermion and  the boson with $Q_B=Q_F-Q_e$ with $B=h,V$. Namely, the $a_{e_a}$ formulas  are   
\begin{align}
	a_{e_a}(h) =& \frac{Q_hm_a }{16\pi^2}  \int_0^1 dx \times \frac{x(x-1) \left[ 2\mathrm{Re}[g^{RL}] m_F + (g^{LL} +g^{RR})m_ax\right]}{(1-x) m_F^2 + x\left[ m_h^2 +m_a^2(x-1)\right]}
	\crn& +\frac{Q_Fm_a }{16\pi^2}\int_0^1 dx \times \frac{x^2 \left[ -2g^{RL}[g^{RL}] m_F + (g^{LL} +g^{RR}) m_a (x-1) \right]}{(1-x) m_h^2 + x\left[ m_F^2 +m_a^2(x-1)\right]},
	\label{eq_amh}	
	\\ a_{e_a}(V) =& -\frac{Q_Vm_a }{16\pi^2 m_V^2}\int_0^1 dx \times \left[  \frac{\mathrm{Re}[g^{RL}]m_F\left[m_F^2(x-1) + m_V^2 x(6x-1) +m_a^2 x(3-5x+2x^2)\right]}{(1-x) m_F^2 + x\left[ m_V^2 +m_a^2(x-1)\right]} \right.
	\crn&\left. - \frac{m_a(g^{LL} +g^{RR})\left[m_F^2(2-3x +x^2) + m_V^2 2x(x+1) +m_a^2 x(x-1)\right]}{(1-x) m_F^2 + x\left[ m_V^2 +m_a^2(x-1)\right]} \right] 
	\crn &+  \frac{Q_Fm_a }{16\pi^2 m_V^2}\int_0^1 dx\left[  \frac{2g^{RL}[g^{RL}]m_Fx\left[m_F^2x - 4m_V^2 (1-x) +m_a^2 x(2x -1 )\right]}{(1-x) m_V^2 + x\left[ m_F^2 +m_a^2(x-1)\right]} \right. 
	\crn &\left.+ \frac{(g^{LL} +g^{RR}) m_a x\left[m_F^2 x(1+x) +2 m_V^2 (2 -3x +x^2) +m_a^2 x(x-1)\right]}{(1-x) m_V^2 + x\left[ m_F^2 +m_a^2(x-1)\right]}  \right], \label{eq_amV}	
\end{align}
where $g^{RL}=g^{R*}_{a,FB}g^{L}_{a,FB}$, $g^{LL}=g^{L*}_{a,FB}g^{L}_{a,FB}$, and $g^{RR}=g^{R*}_{a,FB}g^{R}_{a,FB}$ with $B=h,V$. The coupling identifications are  $ g^R_{a,Fh}= g^{aa}_{s k} +g^{aa}_{pk}$ and  $g^R_{a,Fh}= g^{aa}_{sk} -g^{aa}_{pk}$ for $k=1,2,3$ relating to neutral, singly, and doubly charged Higgs bosons. Similarly to the gauge bosons, $ g^R_{a,FV}= g^{aa}_{v k} +g^{aa}_{ak}$ and  $g^R_{a,FV}= g^{aa}_{vk} -g^{aa}_{ak}$ for $Q_{V}=1,0,-1,2$ corresponding $k=1,2,3,4$.  
The two formulas  \eqref{eq_amh} and \eqref{eq_amV} are derived by inserting the  PV functions given in appendix \ref{app_PVLT} in the limit $p_1^2=p_2^2=m_a^2$ into  $C_{(ab)L,R}$. We have checked that our results are consistent with all  $HFF$, $FHH$, and $VFF$ contributions relating to  the diagrams (1), (2), and (6) in Fig. \ref{fig_eab}, respectively. For the one-loop $FVV$ contributions  arising from the  diagram (5), there is  a difference between  our result and that in Ref. \cite{Lindner:2016bgg}, namely 
$$ \delta (a_{e_a})(FVV)=  \frac{Q_Vm_a m_F}{16\pi^2 m_V^2} (|g^{aa}_{vk}|^2 - |g^{aa}_{ak}|^2)\int_0^1 dx(2x+1) = \frac{Q_Vm_a m_F}{8\pi^2 m_V^2} (|g^{aa}_{vk}|^2 -|g^{aa}_{ak}|^2).  $$    
 It shows that the two results are consistent  if  $g^{aa}_{vk}=\pm g^{aa}_{ak}$,i.e., $g^L_{a,FB}g^R_{a,FB}=0$, which appears in many BSM such as the SM, 3-3-1 models,... We also see that the  $FVV$ contribution to $a_{e_a}$ of the doubly gauge boson given in Ref. \cite{Lindner:2016bgg} has an opposite sign with our result.

We note that our results are also valid as the exact solutions for studying the AMM and $e_b\to e_a\gamma$ decay in BSM  consisting of very light bosons $m_B \ll m^2_{a},m^2_{b}$ such as an   axion-like  particle (ALP) \cite{Bauer:2019gfk, Cornella:2019uxs}, or a new scalar singlet   \cite{Liu:2018xkx}.

\section{\label{sec_conclusion} Conclusion}
 Using the unitary gauge,  we confirm the exact results of analytic formulas in terms of PV functions  for one-loop contributions to the cLFV decay rates  $e_b\to e_a\gamma$ given in  Ref. \cite{Lavoura:2003xp}, which are also applicable to compute the AMM of charged leptons. These results are  consistent with those given in Ref. \cite{Crivellin:2018qmi} in the limit of heavy bosons $m_B\gg m_a,m_b$.   The general expressions in terms of PV-functions are very convenient to change into available forms. Our calculations here have many new features as follows.  Our calculation is independent of the Goldstone boson couplings of the new gauge bosons. The Ward Identity of the external photon allows only the couplings of a photon with two identical physical particles, as  given in table \ref{t_AXX}. At tree-level, the  $ASV$ couplings do not satisfy the WI if $\varepsilon_v.p_{0} \neq 0$, where $\varepsilon_v$ and $p_0$ are the polarization of gauge boson $V$ and the external momentum of the photon, respectively. The one-loop $FSV$ contributions arising from this vertex type to cLFV amplitudes and AMM  do violate the WI. Therefore, the results given in Ref. \cite{Lavoura:2003xp, Crivellin:2018qmi} are valid in all renormalizable BSM respecting the WI. They are still applied for other similar decays of quarks $q\to q'\gamma$. The  photon-scalar-vector $ASV$ vertex does not appear in BSM satisfying the WI. Our conclusion  is very useful for constructing loop calculations relating to photon couplings, where only the vertex types listed in Table \ref{t_AXX} are valid. 

\section*{Acknowledgments}
The authors thank the referee for suggesting an open question about the existence of the $S$-$V$-$\gamma$ couplings in the UV models, which we will solve more generally in our future work. This research is funded by the Vietnam National Foundation for Science and Technology Development (NAFOSTED) under the grant number 103.01-2019.387. 

\appendix 
\section{ \label{app_PVLT} PV functions for one loop contributions defined by LoopTools}
\subsection{General notations}
The PV-functions used here were listed in Ref.~\cite{Hue:2017lak}, namely
\begin{align}
	&A_{0}(m^2)=\frac{(2\pi\mu)^{4-d}}{i\pi^2}\int \frac{d^d k}{k^2 -m^2 +i\delta},\crn
	&B_{\{0,\mu \}}(p^2_i,M^2_1,M^2_2) =\frac{(2\pi\mu)^{4-d}}{i\pi^2}\int \frac{d^d
		k \times\{1, k_{\mu}, k_{\mu}k_{\nu}\}}{D_0 D_i}, \; i=1,2, \crn
	&C_{\{0, \mu, \mu\nu\}}=\frac{(2\pi\mu)^{4-d}}{i\pi^2}\int \frac{d^d k \{ 1,k_\mu,k_\mu k_\nu\}}{D_0 D_1 D_2 },
	\crn &B_{\mu }(p^2_i,M^2_1,M^2_2) = \left( -p_{i\mu}\right) B^{(i)}_1, 
	\crn &C_\mu = \left( -p_{1\mu}\right)C_1 +\left( -p_{2\mu}\right)C_2,\crn
	&C_{\mu\nu} = g_{\mu\nu}C_{00} + p_{1\mu}p_{1\nu}C_{11} +p_{2\mu}p_{2\nu}C_{22}+ (p_{1\mu}p_{2\nu}+p_{2\mu}p_{1\nu})C_{12},
	\label{ABC_def}
\end{align}
where $D_0\equiv k^2-M_1^2  +i\delta$, $D_i\equiv (k -p_i)^2-M_2^2 + i\delta$, $C_{0,\mu,\mu\nu}=C_{0,\mu,\mu\nu}(p_1^2, 0,p_2^2; M_1^2,M_2^2,M_2^2)$,  $\mu$ is an arbitrary mass parameter 
introduced via dimensional regularization \cite{tHooft:1972tcz}. In this work, we discuss only the case of external photon $q^2=(p_2 -p_1)^2=0$. The scalar functions $A_0,B_0, B^{(i)}_1,C_0,C_{00},C_{i}$, and $C_{ij}$ ($i,j=1,2$) are well-known PV functions, which are consistent with those defined by LoopTools \cite{Hahn:1998yk}. The well-known relations are:
\begin{align}
	\label{eq_Bi}	
	& B^{(i)}_0 \equiv B^{(i)}_0(p_i^2; M_1^2, M_2^2)=B^{(i)}_0(p_i^2; M_2^2, M_1^2),
	\crn &B^{(i)}_1 \equiv B^{(i)}_1(p_i^2; M_1^2, M_2^2)= -\frac{1}{2p_i^2}\left[ A_0(M_2^2) -A_0(M_1^2) + f_i B^{(i)}_0\right],
\end{align}
where $f_i=p_i^2 +M_2^2 -M_1^2$. Depending on the  particle exchanges in Feynman diagrams, the $B^{(i)}_1$-function given in Eq. \eqref{eq_Bi} is denoted more precisely as follows:
\begin{align}
\label{eq_Bix}	
 B^{(i)f}_1\equiv B^{(i)}_1(p_i^2; m_F^2,m_B^2), \; B^{(i)v}_1\equiv B^{(i)}_1(p_i^2; m_V^2,m_F^2), \;  B^{(i)h}_1\equiv B^{(i)}_1(p_i^2; m_h^2,m_F^2). 
\end{align}
The scalar functions $A_0$, $B_0$, and  $C_0$ can be calculated using the techniques of \cite{tHooft:1978jhc}.  Other PV functions needed in this work are 
\begin{align}
	B_{0,\mu, \mu\nu}(M_2)= \frac{(2\pi\mu)^{4-d}}{i\pi^2}\int \frac{d^d k~\{1,\; k_{\mu},\; k_{\mu} k_{\nu} \}}{D_1D_2}.
\end{align}
For simplicity, we define the following notations appearing in many important formulas:
\begin{align}
\label{eq_Xidef}
X_0&\equiv C_0 +C_1 +C_2,
\crn   X_1&\equiv C_{11} +C_{12} +C_1, 
\crn   X_2&\equiv C_{12} +C_{22} +C_2,
\crn   X_3&\equiv C_1 +C_2=X_0-C_0,
\crn   X_{012}&\equiv X_0+X_1 +X_2, \; X_{ij}= X_i+X_j.
\end{align}
Depending on the form of the PV-functions, we have
\begin{align}
\label{eq_Xix}
X^f_i& =X_{i}(m_F^2,m_B^2,m_B^2), \; X^h_i \sim X_{i}(m_B^2,m_F^2,m_F^2), \; X^v_i \sim X_{i}(m_V^2,m_F^2,m_F^2)
\end{align}
 corresponding  to the diagram types of  $FBB$, $HFF$, and $VFF$ with $B=h,V$.

\subsection{ $p_1^2\neq p_2^2 \neq0$ and $p_1^2,p_2^2 \neq0$.}

From the definitions of PV-functions given in Eq. \eqref{ABC_def}, it can be proved that: 
\begin{align}
& B^{(0)}_0 \equiv B_0^{(0)}(M_2)\equiv B_0(0; M_2,M_2)=C_{UV} -\ln(M_2^2) +\mathcal{O}(\epsilon) ,
\crn &A_0(M)=M^2\left[ B^{(0)}_0(M) +1\right],\label{eq_a0f}\\
& B^{\mu}(M_2)= \frac{1}{2} B^{(0)}_0(p^{\mu}_1+p^{\mu}_2), \label{eq_Bmunu} \\
&B^{\mu\nu}(M_2)=\frac{g^{\mu\nu}}{2} M^2_2 \left[B^{(0)}_0+1\right]+\frac{1}{6}B^{(0)}_0\left(2p_1^{\mu}p_1^{\nu}+p_1^{\mu}p_2^{\nu}+ p_2^{\mu}p_1^{\nu}+2 p_2^{\mu}p_2^{\nu}\right),
\crn& 	C_{00}=\frac{1}{4}\left[ 2 M_2^2 C_0 + \frac{\left( M_2^2 -M_1^2 +m_a^2\right) B^{(1)}_0 - \left( M_2^2 -M_1^2 +m_b^2\right) B^{(2)}_0}{m_a^2 -m_b^2} +1\right], \label{sbmunu}
\end{align}
where $C_{UV} $  is defined as  the divergent part of the PV functions when $D\to 4$, $C_{UV}=1/\epsilon-\gamma_E + \log(4\pi\mu^2)$ with $\gamma_E$ being Euler's constant and $D=4 -2\epsilon$.  It is well-known that the  PV-functions having non-zero divergent parts are:
\begin{align}
	\label{eq_CuvFPV}
	\mathrm{div}\left[B^{(0)}_0 \right] &=\mathrm{div}\left[B^{(1)}_0 \right] =\mathrm{div}\left[B^{(2)}_0 \right]= -2\mathrm{div}\left[B^{(1)}_1 \right] = -2\mathrm{div}\left[B^{(2)}_1 \right]=4\mathrm{div}\left[C_{00} \right] =C_{UV},
	\crn \mathrm{div}\left[A_0(M) \right] &=M^2 C_{UV}. 
\end{align}
As mentioned in Ref. \cite{Hue:2017lak}, we can derive all formulas of   $C_i$, and $C_{ij}$ as functions of $A_0$, $B^{(i)}_0$, and $C_0$ consistent with Ref. \cite{Hue:2017lak}, using the following  relations: 
\begin{align}
\label{eq_PVrelation0}
&2 m^2_a C_1 + (m_a^2+m_b^2)C_2 = - f_a C_0 -B^{(0)}_0 +B^{(2)}_0,
\crn &   (m_a^2+m_b^2)C_1 + 2m^2_b C_2= - f_b  C_0 -B^{(0)}_0 +B^{(1)}_0,
\crn &  2C_{00}+  2 m^2_a C_{11} + (m_a^2+m_b^2)C_{12} =\frac{1}{2} B^{(0)}_0 - f_a C_1,
\crn &    2 m^2_a C_{12} + (m_a^2+m_b^2)C_{22} =\frac{1}{2} B^{(0)}_0  + B^{(2)}_1 - f_a C_2
\crn &  2C_{00}+ (m_a^2+m_b^2)C_{12} +  2 m^2_b  C_{22} =\frac{1}{2} B^{(0)}_0 - f_b C_2,
\crn &     (m_a^2+m_b^2)C_{11} + 2 m^2_b C_{12} =\frac{1}{2} B^{(0)}_0 +B^{(1)}_1 - f_b C_1,
\crn & 4C_{00} -\frac{1}{2}  +m_a^2 C_{11} +(m_a^2 +m_b^2)C_{12} +m_b^2C_{22} = B^{(0)}_0 +M_1^2 C_0, 
\end{align}
where $f_{a,b}= M_2^2 -M_1^2+m_{a,b}^2$, and $C_{12}=C_{21}$ are used.   In this work, we need just combinations of these PV-functions  for our immediate steps. In particular, we can prove that:
\begin{align}
	\label{eq_Xi}	
	X_{0} &= -\frac{B^{(1)}_0 -B^{(2)}_0}{m_a^2 -m_b^2},
	\crn  X_{12}&= -\frac{B^{(1)}_1 -B^{(2)}_1}{m_a^2 -m_b^2}
	 \crn& = \frac{A_0(M_1^2) -A_0(M_2^2) }{2 m_a^2 m_b^2} +  \frac{(M_1^2 -M_2^2)}{2 (m_a^2 -m_b^2)}\left( \frac{B^{(1)}_0}{m_a^2} -\frac{B^{(2)}_0}{m_b^2}\right) - \frac{1}{2}X_0,
	 %
	%
	\crn m_a^2 B^{(1)}_1 -m_b^2 B^{(2)}_1 
	& =-\frac{1}{2} \left[ \left( m_a^2 +M_2^2 -M_1^2\right) B^{(1)}_0 - \left( m_b^2 +M_2^2 -M_1^2\right) B^{(2)}_0\right],
	\crn \mathbf{b}_1&\equiv  \frac{m_a^2 B^{(1)}_1 -m_b^2 B^{(2)}_1 }{(m_a^2 -m_b^2)} = -(2 C_{00} +m_a^2 X_1 +m_b^2 X_2),
	\crn(2-d)C_{00} +M_2^2C_0&= -2 C_{00}+ \frac{1}{2}+M_2^2 C_0
		\crn & =-\frac{ \left( m_a^2 +M_1^2 -M_2^2\right) B^{(1)}_0 - \left( m_b^2 +M_1^2 -M_2^2\right) B^{(2)}_0}{2(m_a^2-m_b^2)} 
		\crn&=\mathbf{b}_1+(M_2^2-M_1^2) X_0,
\end{align}
where $A_0(M_2^2) =M_2^2 (B^{(0)}_0 +1)$ and $A_0(M_1^2) =M_1^2 (B^{(0)}_0 +1 +\ln (M_2^2/M_1^2))$. 

It was proved previously, for example  \cite{Hue:2017lak},  that
\begin{align}
	\label{eq_bc0}
	&B_{0}(p^2; M^2_1,M^2_2)=	B_{0}(p^2; M^2_2, M^2_1) =C_{UV} - \ln(M^2_2) + 2 - \sum_{\sigma=\pm}(1-\fr{1}{x_\sigma})\ln\left(1-x_\sigma\right),\crn
	&C_{0}(m_a^2,0,m_b^2 ; M^2_1,M^2_2,M^2_2) = -\frac{1}{m_a^2 -m_b^2}\sum_{\sigma=\pm} \left[ \mathrm{Li}_2(y_{a \sigma}) -\mathrm{Li}_2(y_{b \sigma}) \right],
\end{align}
where $p=p_a,p_b$;  and 
\begin{align}
\label{eq_xi}
x_{\pm}&= \frac{1}{2 M_2^2}\left[
 (M_2^2 -M_1^2 +p^2) \pm \sqrt{(M_2^2 -M_1^2 +p^2)^2 -4 M_2^2 p^2}\right],
\crn y_{a\pm}&= \frac{1}{2 M_2^2}\left[
(M_2^2 -M_1^2 +m^2_a) \pm \Lambda\right],
\crn y_{b\pm} &=x_{a\pm}\left[ b\to a\right]
\end{align}
with  $\Lambda =(M_1^4 +M_2^4 + m_a^4 - 2 M_1^2 M_2^2 -2 M_1^2 m_a^2 -2M_2^2 m_a^2)^{1/2}$. 
The above formula of $C_0$ is also consistent with that introduced in loop-induced decay amplitude of $h\to Z \gamma$ \cite{Djouadi:1996yq}.   
\subsection{$m_a^2= p_a^2=p_b^2\neq0$}
Formulas for AMM in Ref. \cite{Yu:2021suw} require that analytic formulas of PV functions with $m_b=m_a$. It seems that the results of PV-functions listed in Ref. \cite{Hue:2017lak} are not valid. But the limit $m_b=m_a$ can be derived mathematically.  For example, the result of $C_0$ given in Eq. \eqref{eq_bc0} leads to a consequence that  
\begin{align}
C_{0}(m_a^2,0,m_a^2; M^2_1,M^2_2,M^2_2)&= \lim_{m_b\to m_a} C_{0}(m_a^2,0,m_b^2; M^2_1,M^2_2,M^2_2)
\crn& =- \frac{\partial}{\partial (m_a^2)}\sum_{\sigma=\pm} \mathrm{Li}_2(y_{a \sigma})  
= \sum_{\sigma=\pm} \frac{y'_{a \sigma} \ln(1-y_{a \sigma})}{y_{a \sigma}} 
\crn& =\sum_{\sigma=\pm} \frac{\ln(1-y_{a\sigma})}{2M_2^2y_{a\sigma}}\times \left[ 1-\sigma \times  \frac{M_1^2 +M_2^2 -m_a^2}{\Lambda} \right] ,
\end{align}
where   $ f'\equiv \partial f/(\partial m_a^2)$ denotes a well-known derivative notation.  In addition, $B^{(1)}_{0}=B^{(2)}_{0}$ and $B^{(1)}_{1}=B^{(2)}_{1}$ are automatically satisfied. Many formulas containing $(m_a^2 -m_b^2)$ in the denominators corresponding a derivative in the limit $m_a \to m_b$:
 \begin{align}
 	\label{eq_mab}
 	X_0&= -B^{(1)\prime}_0=\sum_{\sigma=\pm} \frac{y'_{a \sigma} \left[y_{a \sigma}+ \ln(1-y_{a \sigma})\right]}{y_{a \sigma}^2 } ,
 	\crn X_{12}&= -B^{(1) \prime}_1. 
 %
 %
 \dots, 
 \end{align}
 In this way, we can confirm all results introduced in Ref. \cite{Yu:2021suw}. 
There is another way to calculate form factors, using the Feynman trick:
\begin{align}
	 \frac{1}{D_0D_1D_2}&= \Gamma(3)\int^1_0\frac{dx~dy~dz~\delta(1-x-y-z)}{D^3},\label{feytrick1}
\end{align}
where 
\begin{align}
D&= \left[ k-(yp_1+zp_2)\right]^2 -M^2+i\delta ,
\crn M^2&=  y(y+z-1) p_1^2 +z(y+z-1) p_2^2 +xM_1^2 +(1-x) M_2^2.  \label{d1}	
\end{align}
With $M_0^2= (p_2^2 -p_1^2) xy -x(1-x)p_2^2 +xM_1^2 +(1-x)M_2^2$, the PV functions are:
\begin{align}
\label{eq_PVint}
C_{\{0,1,2,11,22,12\}} &= -\int_0^1dx\int_0^{1-x} \frac{ dy \left\{ 1,-y, -(1-x-y),  y^2, (1-x-y)y, (1-x-y)^2 \right\}}{M_0^2},
\crn X_{0,1,2,3} &=- \int_0^1dx\int_0^{1-x}  \frac{ dy\times \{ x, -xy, -x(1 -x -y) , -(1-x)\}}{M_0^2}. 
\end{align} 
The expressions of $X_{i}$ in Eq. \eqref{eq_PVint} are very convenient for the case of $(g-2)$ anomaly, where $p_1^2=p_2^2=m_a^2$ results in $M_0^2= -x(1-x) m_a^2 +xM_1^2 +(1-x)M_2^2$, which is independent with $y$.  Consequently, the 
\begin{align}
\label{eq_Xia}
X_{0,1,2,3} &=- \int_0^1dx \frac{  \{ x(1-x), -x(1-x)^2/2,  -x (1 -x)^2/2, -(1-x)^2\}}{ M_0^2}
\crn&=- \int_0^1dx \frac{  \{ x(1-x), -(1-x)x^2/2,  -(1-) x^2/2, -x^2\}}{ M_0^2},
\end{align}
Formulas of Eq. \eqref{eq_Xia} are enough to check the consistency between our results   with those  of $(g-2)$ anomalies and cLFV amplitudes  mentioned in ref. \cite{Lindner:2016bgg}. 
Using the second line of Eq. \eqref{eq_Xia}, we can write the general formulas of $a_{\mu}$ as shown in Eqs. \eqref{eq_amh} and \eqref{eq_amV}.

Indeed, all integrals in Eqs. \eqref{eq_amh} and \eqref{eq_amV} can be solved analytically. Starting from the general formulas of $M_0^2$ as  functions of $x$: $M_0^2(x)= m_a^2(x-x_{+})(x- x_{-})$ corresponding to the two solutions  $x_{\pm}$. All numerators in  Eqs. \eqref{eq_amh} and \eqref{eq_amV} are always written in the following forms:
\begin{align}
ax^2 +bx^2+c=a_1 M_0^2 +b_1 \frac{dM_0^2}{dx} +c_1. 	
\end{align}
The consequence is
\begin{align}
	\int_{0}^{1}dx\times\frac{ax^2 +bx^2+c}{M_0^2}=a_1  +b_1 \ln \frac{M_1^2}{M_2^2}  +\frac{c_1}{\sqrt{\Lambda}}\ln\left[\frac{(1-x_-)x_+}{(1-x_+)x_-}\right]. 	
\end{align}
The result in this way must be consistent with those discussed in Ref. \cite{Yu:2021suw}, hence we do not show precisely here. 

\subsection{$p_a^2=p_b^2=0$}
Results for the case of $p^2_a = p^2_b = 0$ were provided in Ref. \cite{Lavoura:2003xp}, namely \begin{align}
	\label{Cformula}
	C_0&=a=\frac{M_1^2-M_2^2 +M_1^2\ln\left[\frac{M_2^2}{M_1^2}\right]}{(M_1^2 -M_2^2)^2},
	\crn C_1&=C_2=c= -\frac{3 M_1^4 -4 M_1^2M_2^2 +M_2^4 +2 M_1^4 \ln\left[\frac{M_2^2}{M_1^2}\right]}{4(M_1^2-M_2^2)^3},
	\crn C_{11}&=C_{22}=2 C_{12}=d= \frac{11 M_1^6 -18M_1^4M_2^2 +9M_1^2M_2^4 -2M_2^6+6M_1^6 \ln\left[\frac{M_2^2}{M_1^2}\right]}{18(M_1^2-M_2^2)^4}.
\end{align}
This approximate formulas of PV functions give  results  consistent with those given in Ref. \cite{Crivellin:2018qmi},
namely 
\begin{align}
	&f_h(x)=2\tilde{g}_h(x)=\dfrac{x^2-1-2x\log x}{4(x-1)^3},
	\nonumber \\
	&g_h(x)=\dfrac{x-1-\log x}{2(x-1)^2},
	\nonumber \\
	&\tilde{f}_h(x)=\dfrac{2x^3+3x^2-6x+1-6x^2\log x}{24(x-1)^4}, 	\label{eq_fgx}\\
	&f_V(x)=\dfrac{x^3-12x^2+15x-4+6x^2\log x}{4(x-1)^3},
	\nonumber \\
	&g_V(x)=\dfrac{x^2-5x+4+3x \log x }{2(x-1)^2},
	\nonumber \\
	&\tilde{f}_{V}(x)=\dfrac{-4x^4+49x^3-78x^2+43x-10-18x^3 \log x}{24(x-1)^4},
	\nonumber 
	\\ &\tilde{g}_{V}(x)=\dfrac{-3(x^3-6x^2+7x-2+2x^2\log x)}{8(x-1)^3},
	\nonumber 
\end{align}
where $x\equiv m_F^2/m_B^2$. The diagrams $FBB$ and $BFF$ correspond to different identifications that $\{M_1,M_2\}=\{ m_F,m_B\}$ or and   $\{M_1,M_2\}=\{ m_B,m_F\}$.

\section{ \label{app_special} Notations in Ref. \cite{Lavoura:2003xp}}

Here we give a brief review of  the approach of Ref. \cite{Lavoura:2003xp}. Apart from the general couplings of physical Higgs and gauge bosons given in Eqs. \eqref{eq_LFh}  and  \eqref{eq_LFV}, the photon couplings were assumed to be the standard forms given in table \ref{t_AXX}. Furthermore, the couplings of the Goldstone boson $G_V$ corresponding to $V$ are assumed to be the following forms:
\begin{align}
	\label{eq_Gvcoupling}
	\mathcal{L}_{G_V}= & \left\{ G_V \frac{i}{m_V} \overline{F} \left[ \left(g^R_{a,FV} m_a - g^L_{a,FV} m_F\right) P_L + \left(g^L_{a,FV} m_a - g^R_{a,FV} m_F\right) P_R \right] e_a +\mathrm{h.c.} \right\}
	\crn&+ e Q_Vm_VA_{\mu}V^{*\mu}G_V - e Q_Vm_VA_{\mu}V^{\mu}G^*_V.
\end{align} 
The above assumptions of the $G_V$ couplings are necessary for the calculation of one-loop gauge contributions that were done in the 't Hoof Feynman gauge. These final results introduced in Ref. \cite{Lavoura:2003xp} were the sum of all diagrams consisting of  gauge and Goldstone boson exchanges.  
Corresponding to the two one-loop diagram classes  $FVV$ and $VFF$, we have the following equivalence between two classes of  notations 
\begin{align*}
\left\{ a,c_1,c_2,d_1,d_2,f,g\right\}&\equiv \left\{ C_0, C_2,C_1,C_{22},C_{11}, C_{12}, C_{00}\right\}^{B},
\crn \left\{ \bar{a}, -\bar{c}_1,-\bar{c}_2, \bar{d}_1, \bar{d}_2,\bar{f}, \bar{g}\right\}&\equiv \left\{ C_0, C_2, C_1,C_{11},C_{22}, C_{12}, C_{00}\right\}^{f}, 
\end{align*}
where  $B=h,V$ are gauge bosons in the loop. In addition,  the different notations in the definitions of the one-loop integrals given in  Eq. \eqref{ABC_def}, we have $\{m_1, m_2\}\equiv \{m_b,m_a\}$  while $\{ p_1, p_2\}\equiv \{-p_2,-p_1\}$ and  $\{ p_1,p_2\}\equiv \{p_2,p_1\}$ for the diagrams $VFF$ and $FVV$ respectively. The couplings in the Yukawa Lagrangian of physical bosons are $L_1\equiv g_b^L$, $R_1\equiv g_b^R$, $L_2\equiv g_a^L$, and $R_2\equiv g^a_R$, which result in the following equivalences: $\lambda\equiv g^{L*}_a g^{L}_b=g^{LL}$, $\rho\equiv g^{R*}_ag^{R}_b= g^{RR}$, $\zeta \equiv g^{L*}_a g^{R}_b =g^{LR}$, and $v \equiv g^{R*}_a g^{L}_b =g^{RL}$.  As a result, we can identify that:
\begin{align}
\label{eq_Lavoura2003}
k_1 &=m_b X_2^B,\; k_2=m_a X_1^B, \; 
  k_3 =m_F(c_1 +c_2)=m_FX_3^B,
\crn  	\bar{k}_1 &= m_b X_2^f,\; \bar{k}_2  =m_b X_1^f, \;
%
 k_3  = -m_F X_3^f. 
\end{align}
 For a gauge boson $B_{\mu}$, the one-loop form factors  relate to the following  notations: 
\begin{align}
\label{eq_yi}
y_1 &= m_b \left[ 2X_{02}^f + \frac{m_F^2(X^f_2-X^f_3) +m_a^2X^f_1}{m_B^2}\right], \; 
 y_2  = m_a \left[ 2X_{01}^f + \frac{m^2_F(X_1^f -X_3^f) +m_b^2X_2}{m^2_B}  \right],	
\crn y_3 
&= m_F\left[ -4X_0^f + \frac{m^2_F X_3^f +m^2_aX_1^f  +m^2_bX^f_2}{m^2_B}\right],
 \; y_4 = -\frac{m_am_b m_F (X_{12}^f- X_3^f)}{m^2_B},
\end{align}
and 
\begin{align}
\label{eq_byi}
	\bar{y}_1 & = m_b\left[ 2(X_2^f -X_3^f) + \frac{m^2_FX_{02}^f +m_a^2 X_1^f}{m^2_B}\right],
	\; \bar{y}_2  = m_a \left[ 2(X_1^f -X_3^f) +\frac{m^2_FX_{01}^f +m_b^2 X_2^f}{m^2_B} \right],
	\crn \bar{y}_3  &=m_F \left[ 4X_3^f + \frac{-m^2_FX_0 -m_a^2X_1 -m_b^2X_2}{m^2_B} \right], 
	 \; \bar{y}_4  =- \frac{m_a m_bm_F}{m^2_B}X_{012}. 
\end{align}

\section{ \label{app_detailedStep} Important steps to derive $C_{(ab)L,R}$ and  $D_{(ab)L,R}$ by hand}

\begin{figure}[ht] 
	\centering
	\includegraphics[scale=0.8]{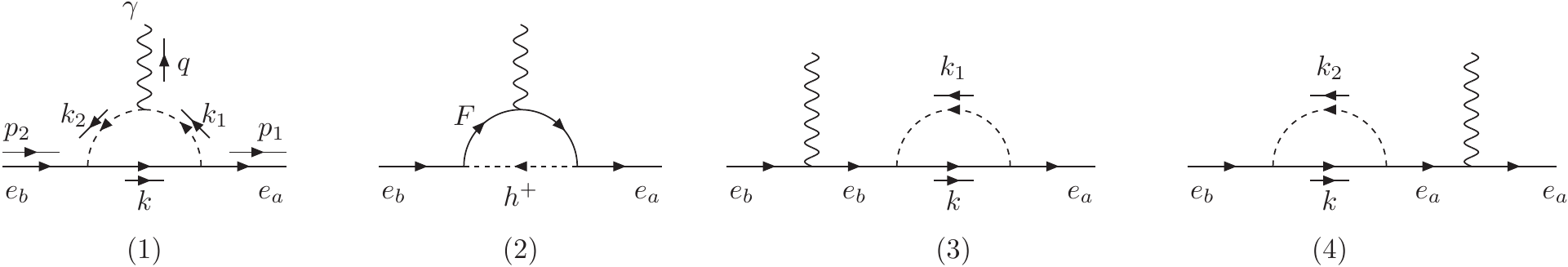}
	\caption{Momneta notations to derive the one-loop contributions} \label{fig:FH}
\end{figure}

The notations for calculating the amplitude corresponding to all diagrams of both Higgs and gauge boson exchanges  in Fig. \ref{fig_eab} are shown in Fig. \ref{fig:FH}.  
   All diagrams in the same class will have the same conventions of external momenta and propagators. There are three classes of diagrams: i) The first class consists of four diagrams (1), (2), (5), and (6) in Fig. \ref{fig_eab}, and  the two diagrams (1) and (2) in Fig. \ref{fig:FH}; ii) the second class consists of three diagrams: (3) and (7) in Fig. \ref{fig_eab}, and (3) in Fig. \ref{fig:FH}; iii) the last class consists of the remaining diagrams in the two Figs. \ref{fig_eab} and \ref{fig:FH}.  Although all the internal momenta  have opposite signs with those denoted following LoopTools, the PV-functions  are defined with the same values.  The relations relevant to momenta are:
 \begin{align}
 k_i &=k-p_i,\; p_1^2=m^2_{a}, \;  p_2=q +p_1,\;	p_2^2=m^2_{b}, \; q^2=0, \crn
q.\varepsilon^*&= 0, \; p_1.\varepsilon^*=p_2.\varepsilon^*,\;  \end{align}
Only four diagrams (1), (2), (5), and (6) in Fig. \ref{fig_eab} give non-zero contributions to $C_{(ab)L,R}$, hence we firstly derive $C_{(ab)L,R}$ as the factors of $(2p_1.\varepsilon^*)$ in the amplitudes arising from these diagrams. For convenience in detailed calculations, we use simple notations for all the coupling factors  $ g^{aL,R}_{FB}\to g^{L,R}_{a}$. For integrals containing divergences, we use the regular  dimensional  regularization defined by the following replacement:
$$ \int \frac{d^4k}{(2\pi)^4} \to \frac{i}{16 \pi^2} \times  \frac{(2\pi\mu)^{4-d}}{i\pi^2} \int d^d k \equiv  \int D k. $$
 The final results now are written in terms of the PV functions.  In many intermediate steps, we use many results for products of gamma matrices in the dimension $d$ \cite{Peskin:1995ev}, namely
\begin{align*}
	&\gamma^{\mu}\gamma_{\mu}=d,
	\crn & \gamma^{\mu} \gamma^{\nu}\gamma_{\mu}= (2-d) \gamma^{\nu} \to \gamma^{\mu} \slashed{p}\gamma^{\mu}=(2-d) \slashed{p},
	\crn & \gamma^{\mu} \gamma^{\nu}\gamma^{\rho} \gamma_{\mu}= 4 g^{\nu \rho} +(d-4) \gamma^{\nu}\gamma^{\rho} \to \gamma^{\mu} \slashed{p}_1\slashed{p}_2 \gamma^{\mu}= 4p_1.p_2 +(d-4) \slashed{p}_1\slashed{p}_2,
	\crn & \gamma^{\mu} \gamma^{\nu}\gamma^{\rho}\gamma^{\sigma} \gamma_{\mu}= -2 \gamma^{\sigma} \gamma^{\rho} \gamma^{\nu}   -(d-4) \gamma^{\nu}\gamma^{\rho} \gamma^{\sigma} \to \gamma^{\mu} \slashed{p}_1\slashed{p}_2 \slashed{p}_3 \gamma_{\mu}= -2 \slashed{p}_3 \slashed{p}_2 \slashed{p}_1 - (d-4) \slashed{p}_1\slashed{p}_2 \slashed{p}_3,\dots
\end{align*}
 \subsection{Scalar contributions}
 We list here 8 formulas of amplitudes corresponding to  8 particular diagrams shown in Fig. \ref{fig_eab}. Namely,  for three diagrams (1), (3), and (4) we have 
\begin{align} 
i\mathcal{M}_1 
%
%
= & - eQ_H \int Dk\times  \overline{u_{a}} [{g_{a}^{R}}^{*}P_L + {g_{a}^{L}}^{*}P_R]  \frac{(m_F + \slashed{k})}{D_0D_1 D_2} [g_{b}^{L}P_L + g_{b}^{R}P_R]   u_{b} 
\times (2k_1.\varepsilon^*)
, \label{eq_M1}
\\ i\mathcal{M}_3 =& \frac{-eQ_e}{m^2_a -m^2_b} \int Dk\times  \overline{u_{a}} [{g_{a}^{R}}^{*}P_L + {g_{a}^{L}}^{*}P_R]  \frac{(m_F + \slashed{k})}{D_0D_1 } [g_{b}^{L}P_L + g_{b}^{R}P_R]   (m_b +\slashed{p}_1)\slashed{\varepsilon}^*u_{b},   \label{eq_M3}
\\ i\mathcal{M}_4 =&  \frac{eQ_e}{m^2_a -m^2_b} \int Dk\times  \overline{u_{a}}  \slashed{\varepsilon}^* (m_a +\slashed{p}_2) [{g_{a}^{R}}^{*}P_L + {g_{a}^{L}}^{*}P_R]  \frac{(m_F + \slashed{k})}{D_0D_2} [g_{b}^{L}P_L + g_{b}^{R}P_R] u_{b},  \label{eq_M4}
\end{align}
where $D_0=k^2 -m_F^2$ and $D_i=k_i^2 -m_h^2$.
The amplitude for the diagram (2) is:
\begin{align}
\label{eq_M2}	
 i\mathcal{M}_2 &= -eQ_F \int Dk \times   \overline{u_{a}} [{g_{a}^{R}}^{*}P_L + {g_{a}^{L}}^{*}P_R]  \frac{(m_F - \slashed{k}_1)\slashed{\varepsilon}^{*}  (m_F - \slashed{k}_2)}{D_0D_1D_2}  [g_{b}^{L}P_L + g_{b}^{R}P_R]  u_{b},  
\end{align} 
where $D_0=k^2 -m_h^2$ and $D_i=k_i^2 -m_F^2$.

In the next calculation, we use the following simple notations:
\begin{align}
\label{eq_gLR}
g^{LL} &\equiv g_{a}^{L*}g_{b}^{L}, \;  g^{RR}\equiv g_{a}^{R*}g_{b}^{R},  \; g^{RL}\equiv g_{a}^{R*}g_{b}^{L}, \; g^{LR}\equiv g_{a}^{L*}g_{b}^{R},
\crn A_1&= g_{a}^{L*}g_{b}^{R}P_R + g_{a}^{R*}g_{b}^{L}P_L, \; A_2= g_{a}^{L*}g_{b}^{L}P_L + g_{a}^{R*}g_{b}^{R}P_R,
\end{align}
where $ g_{a}^{L,R}\equiv  g_{a,Fh}^{L,R}$ and $ g_{b}^{L,R}\equiv  g_{b,Fh}^{L,R}$ without any confusions with the gauge boson couplings $ g_{a,FV}^{L,R}$.
It is not hard to write all amplitudes in terms of PV-functions as follows:
\begin{align} 
\mathcal{M}_1  
= &  \frac{-eQ_H}{16\pi^2 } \overline{u_{a}}\left\{ -2p_1.\varepsilon^*  \left[A_1 \right]    m_F  X_0  
+  \left[ 2C^f_{00}\slashed{\varepsilon}^* + \left( X_1^f \slashed{p}_1 + X_2^f \slashed{p}_2 \right) (2p_1.\varepsilon^*) \right]
%
[A_2]    \right\}  u_{b} , \label{eq_M1a}
\\ \mathcal{M}_3 =& \frac{-eQ_e}{16\pi^2 } \times  \frac{\overline{u_{a}} [{g_{a}^{R}}^{*}P_L + {g_{a}^{L}}^{*}P_R]  (m_F B^{(1)}_0 - B^{(1)f}_1 \slashed{p}_1) [g_{b}^{L}P_L + g_{b}^{R}P_R]   (m_b +\slashed{p}_1)\slashed{\varepsilon}^*u_{b}}{(m^2_a -m^2_b)},   \label{eq_M3a}
\\ \mathcal{M}_4 =&  \frac{eQ_e}{16\pi^2} \times \frac{ \overline{u_{a}}  \slashed{\varepsilon}^* (m_a +\slashed{p}_2) [{g_{a}^{R}}^{*}P_L + {g_{a}^{L}}^{*}P_R]  (m_F B^{(2)}_0 - B^{(2)f}_1 \slashed{p}_2) [g_{b}^{L}P_L + g_{b}^{R}P_R] u_{b} }{(m^2_a -m^2_b)} ,  \label{eq_M4a}
\end{align}
and 
\begin{align}
\label{eq_M2a}	
\mathcal{M}_2= & -eQ_F    \int Dk\times  \overline{u_{a}} \left\{ m_F^2 \slashed{\varepsilon}^* + \slashed{k}_1\slashed{\varepsilon}^* \slashed{k}_2 \right\}  
 [A_2]  u_{b} 
\crn & -eQ_F(-1)m_F   \int Dk\times  \overline{u_{a}} \left\{2k.\varepsilon^*  -\slashed{p}_1\slashed{\varepsilon}^* -\slashed{\varepsilon}^*\slashed{p}_2 \right\} [A_1]  u_{b}
\crn =&  \frac{-eQ_F}{16 \pi^2}  \overline{u_{a}} \left\{ \left[ m_F^2 C_0 + (2-d) C_{00}\right]   \slashed{\varepsilon}^*  + (C_{11} +C_1)\slashed{p}_1\slashed{\varepsilon}^* \slashed{p}_1 +   (C_{22} +C_2)\slashed{p}_2 \slashed{\varepsilon}^* \slashed{p}_2 
\right. \crn&  \left.  \qquad  \qquad  +(X_0+C_{12})\slashed{p}_1 \slashed{\varepsilon}^* \slashed{p}_2  + C_{12} \slashed{p}_2 \slashed{\varepsilon}^* \slashed{p}_1  \right\}  
\times [A_2]  u_{b}
\crn & -\frac{eQ_Fm_F}{16 \pi^2}  \overline{u_{a}} \left\{(2p_1.\varepsilon^*) (C_1 +C_2)  +\left( \slashed{p}_1\slashed{\varepsilon}^* +\slashed{\varepsilon}^*\slashed{p}_2\right) C_0  \right\} [A_1]  u_{b}. 
\end{align} 
  The  validation of the WI given in Eq. \eqref{eq_DLR} implies  whether $f^{WI}_{L}=0$ is correct  with:
\begin{align}
\label{eq_fWI}
f^{WI}_{L} \equiv & D_{(ab)L} +m_a C_{(ab)L} +m_b C_{(ab)R}
\crn =& g^{LL} \left[    \frac{ Q_e \left( m_a^2 B^{(1)}_1-   m_b^2 B^{(2)}_1\right)^{f}  }{m_a^2 -m_b^2} -\left(\frac{1}{2} -2C_{00}^h +m_F^2 C_0^h\right)  Q_f 
\right.\crn &\left. \qquad \quad 
 -Q_h\left( m_a^2 X_1  + m_b^2 X_2  +2C_{00}\right)^f  \frac{}{} \right]
\crn & +  g^{RR} m_a m_b \left( \frac{ Q_e \left( B^{(1)}_1-  B^{(2)}_1\right)^{f}  }{m_a^2 -m_b^2} -  Q_f  X_{012}^h - Q_h  X_{12}^f\right).
\end{align}
We have used many formulas  listed in Eqs. \eqref{eq_Bi} and  \eqref{eq_Xi} to show  that 
\begin{align}
\label{eq_Ih}
0&= X_{12}^f +X^h_{12} +X_0 \to X^h_{012}=-X^f_{12},
\crn \mathbf{b}_1^f&= -\left( m_a^2 X_1  + m_b^2 X_2  +2C_{00}\right)^f=  \frac{1}{2} -2C_{00}^h +m_F^2 C_0^h. 
\end{align}
Finally, the electric charge conservation $Q_F=Q_e +Q_h$ must be satisfied so  that Eq. \eqref{eq_fWI} resulting in  $f^{WI}_{L}=0$. On the other word, the WI is valid for only one-loop Higgs contributions arising from the set of four diagrams (1)-(4) in Fig. \ref{fig_eab}.

\subsection{Vector contributions}
To calculate the one-loop contributions from gauge boson exchanges corresponding to Lagrangian \eqref{eq_LFV}, we denote  $ g_{a}^{L,R}\equiv  g_{a,FV}^{L,R}$ and $ g_{b}^{L,R}\equiv  g_{b,FV}^{L,R}$ then use the notations given in Eq. \eqref{eq_gLR}. The amplitudes relevant with gauge boson exchanges are:
\begin{align} 
i\mathcal{M}_5 
=& \int Dk\times  \overline{u_{a}}  i \gamma_\alpha [g_{a}^{L^*}P_L + g_{a}^{R^*}P_R] \frac{i(m_F + \slashed{k})}{D_0}   
i \gamma_\beta[g_{b}^{L}P_L + {g_{b}^{R}}P_R] u_{b}
\nonumber\\& \times 
\frac{-i}{D_1}\left( g^{\alpha \alpha'}-\frac{k_1^\alpha k_1^{\alpha'}}{m^2_V} \right) \left[ -ie Q_V \Gamma_{\mu \alpha' \beta'}\left(-q,k_1,-k_2\right) \varepsilon^{*\mu}\right]\frac{-i}{D_2}\left( g^{\beta \beta'}-\frac{k_2^\beta k_2^{\beta'}}{m^2_V} \right)
\nonumber\\ = &  eQ_V \int \frac{d^4k}{(2\pi)^4} \overline{u_{a}}   \gamma_\alpha [g_{a}^{L^*}P_L + g_{a}^{R^*}P_R] \frac{(m_F + \slashed{k})}{D_0 D_1 D_2}   
\gamma_\beta[{g_{b}^{L}}P_L + {g_{b}^{R}}P_R] u_{b}
\nonumber\\& \times 
\left[ \Gamma_{\mu \alpha' \beta'}\left(-q,k_1,-k_2\right) \varepsilon^{*\mu}\right] \left( g^{\alpha \alpha'}-\frac{k_1^\alpha k_1^{\alpha'}}{m^2_V} \right)\left( g^{\beta \beta'}-\frac{k_2^\beta k_2^{\beta'}}{m^2_V} \right),  \label{eq_M5}
\\ i\mathcal{M}_7 =& \frac{e Q_e}{m_a^2 -m_b^2} \int Dk\times \frac{1}{D_0 D_1} \times \left( g^{\alpha \beta}- \frac{k^{\alpha}_1 k^{\beta}_1}{m_V^2}\right)  
\crn &\times \overline{u_{a}}   \gamma_\alpha [g_{a}^{L^*}P_L + g_{a}^{R^*}P_R] (m_F + \slashed{k})     
\gamma_\beta[{g_{b}^{L}}P_L + {g_{b}^{R}}P_R] \left( m_b + \slashed{p}_1 \right) \slashed{\varepsilon}^*u_{b} ,  \label{eq_M7}
\\ i\mathcal{M}_8 =&- \frac{e Q_e}{m_a^2 -m_b^2} \int Dk\times \frac{1}{D_0 D_2} \times  \left( g^{\alpha \beta}- \frac{k^{\alpha}_2  k^{\beta}_2}{m_V^2}\right) 
\crn &\times \overline{u_{a}} \slashed{\varepsilon}^* \left( m_a + \slashed{p}_2 \right)  \gamma_\alpha [g_{a}^{L^*}P_L + g_{a}^{R^*}P_R] (m_F + \slashed{k})     
\gamma_\beta[{g_{b}^{L}}P_L + {g_{b}^{R}}P_R] u_{b},  \label{eq_M8}
\end{align} 
where $D_0=k^2 -m_F^2$,  $D_i=k_i^2 -m_V^2$, and 
\begin{align}
\label{eq_Gamab}
\Gamma_{\mu \alpha' \beta'}\left(-q,k_1,-k_2\right)= g_{\alpha' \beta'} \left( k_1 +k_2\right)_{\mu}
+ g_{ \beta'\mu} \left(-k_2 +q\right)_{\alpha'} + g_{ \mu \alpha'} \left( -q -k_1 \right)_{\beta'}. 
\end{align}

The amplitude for the diagram (6) is:
\begin{align}
\label{eq_M6}	
i\mathcal{M}_6 = & e Q_F \int Dk\times 
\frac{1}{D_0 D_1D_2} \times \left( g^{\alpha \beta}-\frac{k^\alpha k^\beta}{m^2_V} \right) 
\nonumber \\ &\times  \overline{u_{a}} \gamma_\alpha [g_{a}^{L^*}P_L + g_{a}^{R^*}P_R] (m_F - \slashed{k}_1)  \slashed{\varepsilon}^{*}(m_F - \slashed{k}_2)\gamma_\beta[{g_{b}^{L}}P_L + {g_{b}^{R}}P_R] u_{b},  
\end{align} 
where $D_0=k^2 -m_V^2$ and $D_i=k_i^2 -m_F^2$.

Considering diagram (7), we have:
\begin{align}
\label{eq_M71}
 i\mathcal{M}_7 =& \frac{e Q_e}{m_a^2 -m_b^2} \int Dk\times  \frac{\overline{u_{a}}  \left[ \gamma_\alpha\gamma_\beta m_F \left[A_1\right]  + \gamma_\alpha \slashed{k} \gamma_\beta  \left[ A_2\right]\right] \left( m_b + \slashed{p}_1 \right) \slashed{\varepsilon}^*u_{b} }{D_0 D_1} \times \left( g^{\alpha \beta}- \frac{k^{\alpha}_1 k^{\beta}_1}{m_V^2}\right)  
\crn =& \frac{e Q_e}{m_a^2 -m_b^2} \int Dk\times \frac{1}{D_0 D_1} 
\crn &
\times \overline{u_{a}}  \left[  m_F \left( d  -\frac{k_1^2}{m_V^2}\right) \left[ A_1 \right] 
%
+  \left( (2-d) \slashed{k} -\frac{\slashed{k}_1 \slashed{k} \slashed{k}_1}{m_V^2}\right)  \left[A_2\right]\right] \left( m_b + \slashed{p}_1 \right) \slashed{\varepsilon}^*u_{b}
\crn =&  \frac{ie Q_e}{ 16\pi^2 (m_a^2 -m_b^2)} \overline{u_{a}}  \left\{m_F \left[A_1\right]   \left[  (d-1) B^{(1)}_0  - \frac{A_0(m_F^2)}{m_V^2}\right] 
\right. \crn &\left. \qquad\qquad\qquad  \qquad + m_a  \left[ \left( -(2-d) + \frac{m_F^2 +m_a^2}{m_V^2} \right) B^{(1)}_1   + \frac{A_0(m_V^2) +2m_F^2B^{(1)}_0}{m_V^2}   \right]  \left[A_2\right]  \right\}
\crn & \times \left( m_b + \slashed{p}_1 \right) \slashed{\varepsilon}^*u_{b},
\end{align}
where we have used the following results
\begin{align*}
\slashed{k}_1 \slashed{k} \slashed{k}_1&= \left(D_0 +m_F^2\right) \slashed{k} -2\left(D_0 +m_F^2\right) \slashed{p}_1 +\slashed{p}_1\slashed{k} \slashed{p}_1,
\crn \int \frac{d^4k}{(2\pi)^4} \times \frac{k_{\mu}}{ D_1}&= A_0(m_V^2) p_{1\mu}. 
\end{align*}

Then the one-loop contribution form factors from diagram (7)  are:
\begin{align}
\label{eq_DabLR7}
D_{(ab)L,7}&= \frac{e Q_e}{ 16\pi^2 (m_a^2 -m_b^2)} \left\{ \left(  g^{RL} m_a +  g^{LR} m_b \right)m_F \left[ (d-1) B^{(1)}_0 -\frac{A_0(m_F^2)}{m_V^2}\right]
\right. \crn &+\left. m_a \left( m_a g^{LL} +m_b g^{RR} \right)   \left[ \left(-(2-d) +  \frac{m_F^2 +m_a^2}{m_V^2}\right) B^{(1)}_1 + \frac{A_0(m_V^2) +2m_F^2B^{(1)}_0}{m_V^2} \right] \right\},
\crn D_{(ab)R,7} =&  D_{(ab)L,7}\left[ g^{L}_a \leftrightarrow g^{R}_a, g^{L}_b \leftrightarrow g^{R}_b\right]. 
\end{align}
The same calculation for diagram (8) gives  the following  one-loop contribution form factor:
\begin{align}
\label{eq_DabLR8}
D_{(ab)L,8}&= -\frac{e Q_e}{ 16\pi^2 (m_a^2 -m_b^2)} \left\{ \left(   g^{RL}m_a + g^{LR} m_b\right)m_F \left[ (d-1) B^{(2)}_0 -\frac{A_0(m_F^2)}{m_V^2} \right]
\right. \crn &+\left. m_b \left( m_a g^{RR} + m_b g^{LL}\right)   \left[ \left(-(2-d) +  \frac{m_F^2 +m_b^2}{m_V^2}\right) B^{(2)}_1 +\frac{A_0(m_V^2) +2m_F^2B^{(2)}_0}{m_V^2} \right] \right\},
\crn D_{(ab)R,8}&= D_{(ab)L,8} \left[ g^{L}_a \leftrightarrow g^{R}_a, g^{L}_b \leftrightarrow g^{R}_b\right].
\end{align}
Using $d=4 -2 \epsilon $ and the divergent parts of PV-functions given in Eq. \eqref{eq_CuvFPV}, we get the formulas of $D_{(ab)L,78}$ given in Eq. \eqref{eq_DabLR78}.

\begin{center}
\textbf{{Diagram (5)}}
\end{center}

From the equalities  $q^2=0$, $q.\varepsilon^*=0$, and  $k_1=q+k_2$, it is easy to prove that 
\begin{align}
	\label{eq_Gak4}
	&\left[ \Gamma_{\mu \alpha' \beta'}\left(-q,k_1,-k_2\right) \varepsilon^{*\mu}\right]k_1^\alpha k_1^{\alpha'} k_2^\beta k_2^{\beta'}
	\crn &= k_1^\alpha  k_2^\beta \left\{ (k_1.k_2) \left[(k_1+k_2).\varepsilon^*\right] + (k_2.\varepsilon^*) \left[k_1.(-k_2+q)\right] +  (k_1.\varepsilon^*) \left[k_2.(-q -k_1)\right]
	\right\}
	\crn&\sim   (k_1.k_2) \left[2k_1.\varepsilon^*\right] + (k_1.\varepsilon^*) \left[q^2 -k_2^2\right] +  (k_1.\varepsilon^*) \left[q^2 -k^2_1\right] =0. 
\end{align}
As a result, the amplitude \eqref{eq_M5} is  written as follows:
\begin{align}
	\label{eq_FVV2}
	i\mathcal{M}_5&=  eQ_V \int Dk \frac{\overline{u_a}   \gamma_\alpha \left[ A \right] 	\gamma_\beta u_{b}
	  }{D_0 D_1 D_2} 
	%
	 \left[ \Gamma_{\mu \alpha' \beta'}\left(-q,k_1,-k_2\right) \varepsilon^{*\mu}\right] \left( g^{\alpha \alpha'}g^{\beta \beta'}   -\frac{g^{\beta \beta'} k_1^\alpha k_1^{\alpha'} +g^{\alpha \alpha'} k_2^\beta k_2^{\beta'}}{m^2_V} \right),
\end{align}
where 
\begin{equation}\label{eq_Adef}
A= m_F \left(g_{a}^{L*}g_{b}^{R}P_L + g_{a}^{R*}g_{b}^{L}P_R\right) + \left[A_2\right]\slashed{k}. 	
\end{equation}
The first term in the integrand is 
	\begin{align}
	\label{eq_MA1}
	(1)  =& \overline{u_{a}} \left\{ 4 (k_1 +k_2).\varepsilon^* + (-\slashed{k} +2\slashed{p}_2 -\slashed{p}_1)\slashed{\varepsilon}^* + \slashed{\varepsilon}^* (-\slashed{k} +2\slashed{p}_1 -\slashed{p}_2)\right\} 
	\times m_F\left[A_1\right] u_{b}
	\crn  & +\overline{u_{a}} \left\{(2-d)(2k_1.\varepsilon^*) \slashed{k} +(-\slashed{k}+ 2\slashed{p}_2 -\slashed{p}_1)\slashed{k}\slashed{\varepsilon}^* +\slashed{\varepsilon}^*\slashed{k}(-\slashed{k} +2\slashed{p}_1 -\slashed{p}_2)\right\}
	%
	\times \left[A_2\right]u_{b}
	\crn = &  \overline{u_{a}} \left\{  6k.\varepsilon^*  -3(\slashed{p}_1\slashed{\varepsilon}^*+ \slashed{\varepsilon}^*\slashed{p}_2) \right\} 
	m_F \left[A_1\right]  u_{b}
	\crn  & +\overline{u_{a}} \left\{(2-d)(2k_1.\varepsilon^*) \slashed{k} +( 2\slashed{p}_2 -\slashed{p}_1)\slashed{k}\slashed{\varepsilon}^* +\slashed{\varepsilon}^*\slashed{k}(2\slashed{p}_1 -\slashed{p}_2) -2 k^2 \slashed{\varepsilon}^* \right\}
	%
	\left[A_2\right]u_{b}.
\end{align}
After integrating out, the formula is
\begin{align}
	\label{eq_MA1b}
	(1) =& 	 \overline{u_{a}} \left\{  (2p_1.\varepsilon^*)\times (-3m_F) X_3
	%
	 -3m_FC_0(\slashed{p}_1\slashed{\varepsilon}^*+ \slashed{\varepsilon}^*\slashed{p}_2)  \right\} 
	\left[A_1\right] u_{b}
	\crn  & +\overline{u_{a}} \left\{(2-d)2\varepsilon^{\alpha*}(C_{\alpha \beta} -C_{\beta}p_{1\alpha})\gamma^{\beta}  + C_{\alpha}\left[ ( 2\slashed{p}_2 -\slashed{p}_1) \gamma^{\alpha}\slashed{\varepsilon}^* + \slashed{\varepsilon}^*\gamma^{\alpha}(2\slashed{p}_1 -\slashed{p}_2)\right]
	\right.\crn&\left.  -2 \left( B^{(0)}_0 + m_F^2 C_0\right)\slashed{\varepsilon}^*\right\}  \times \left[A_2\right]u_{b}
	%
	\crn=  & 	 \overline{u_{a}} (-3m_F)\times \left\{  (2p_1.\varepsilon^*)  X_3  +C_0(\slashed{p}_1\slashed{\varepsilon}^* + \slashed{\varepsilon}^*\slashed{p}_2)     \right\} 
\left[A_1\right]	u_{b}
	\crn  & +\overline{u_{a}} \left\{ \slashed{\varepsilon}^* \left[ 2(2-d)C_{00} -2(B^{(0)}_0 +m_F^2 C_0) -(3m_a^2 +2 m_b^2)C_1 -(2m_a^2 +3 m_b^2) C_2\right] 
	\right.\crn&\left. 
	 + \slashed{p}_1\slashed{\varepsilon}^*\slashed{p}_2 (-3X_3)  \right\} \left[A_2\right]u_{b}
	\crn  & +\overline{u_{a}}(2p_1.\varepsilon^*) \left\{  \left[-2 (C_{11} +C_{12})  +C_2\right]
	 \slashed{p}_1 
	+\left[-2 (C_{12} +C_{22}) +C_1 \right]\slashed{p}_2   \right\}
	%
	 \left[A_2\right]u_{b}. 
\end{align}

The second term in  the integrand  is 
\begin{align}
	\label{eq_MA2a}
	& \left(-\frac{1}{m_V^2}\right)^{-1} \times (2) 
	\crn =&\Gamma_{\mu \alpha' \beta'}\left(-q,k_1,-k_2\right) \varepsilon^{*\mu}    \left( g^{\beta \beta'} k_1^\alpha k_1^{\alpha'} +g^{\alpha \alpha'} k_2^\beta k_2^{\beta'}\right)  \times \overline{u_a}   \gamma_\alpha \left[ A \right]    
	\gamma_\beta u_{b}
	\nonumber\\ = & \; \overline{u_a}   \slashed{k}_1\left[ A \right] \left[(k_1.\varepsilon^*)\slashed{k}_2 -\slashed{\varepsilon}^*k_2^2   \right]    
	u_{b} + \overline{u_a}    \left[(k_1.\varepsilon^*)\slashed{k}_1 -\slashed{\varepsilon}^*k_1^2   \right]    
	\left[ A \right]\slashed{k}_2 u_{b} 
	\crn =& \; \overline{u_a}  m_F \left[A_1\right]  \left[ 2(k_1.\varepsilon^*) \slashed{k}_1\slashed{k}_2 - k_2^2\slashed{k}_1\slashed{\varepsilon}^* -  k_1^2\slashed{\varepsilon}^*\slashed{k}_2 \right]    
	u_{b}
	\crn &+ \overline{u_a}   \left[ 2(k_1.\varepsilon^*) \slashed{k}_1\slashed{k} \slashed{k}_2 - k_2^2\slashed{k}_1\slashed{k} \slashed{\varepsilon}^* -  k_1^2\slashed{\varepsilon}^*\slashed{k} \slashed{k}_2 \right]    \left[ A_2\right]
	u_{b}
	%
	\crn =& \; \overline{u_a}  m_F\left[A_1\right]  \left[ \slashed{k}_1\slashed{\varepsilon}^* \slashed{q} \slashed{k}_2 \right] u_{b}
	%
	+ \overline{u_a}  \left[ 2(k_1.\varepsilon^*) \slashed{k}_1\slashed{k} \slashed{k}_2 - k_2^2\slashed{k}_1\slashed{k} \slashed{\varepsilon}^* -  k_1^2\slashed{\varepsilon}^*\slashed{k} \slashed{k}_2 \right]    \left[A_2\right]
	u_{b}. 
\end{align}
The first term in Eq. \eqref{eq_MA2a} gives 
\begin{align}
	\label{eq_sh2l1}
	\slashed{k}_1\slashed{\varepsilon}^* \slashed{q} \slashed{k}_2 & =   \left( \slashed{k} -\slashed{p}_1\right) \slashed{\varepsilon}^* \slashed{q} \left( \slashed{k} -\slashed{p}_2\right) = \slashed{k}  \slashed{\varepsilon}^* \slashed{q}\slashed{k}  -   \slashed{p}_1  \slashed{\varepsilon}^* \slashed{q}\slashed{k} -\slashed{k}  \slashed{\varepsilon}^* \slashed{q}\slashed{p}_2 + \slashed{p}_1  \slashed{\varepsilon}^* \slashed{q}\slashed{p}_2 
	\crn &=  C_{\alpha\beta} \gamma^{\alpha} \slashed{\varepsilon}^* \slashed{q}\gamma^{\beta} -C_{\alpha}  \slashed{p}_1 \slashed{\varepsilon}^* \slashed{q}\gamma^{\alpha} - C_{\alpha}   \gamma^{\alpha}\slashed{\varepsilon}^* \slashed{q} \slashed{p}_2    + C_0 \slashed{p}_1  \slashed{\varepsilon}^* \slashed{q}\slashed{p}_2
	\crn& +   \left( C_{1}\slashed{p}_1
	+C_{2}\slashed{p}_2\right) \slashed{\varepsilon}^* \slashed{q} \slashed{p}_2 	+   \slashed{p}_1 \slashed{\varepsilon}^* \slashed{q} \left( C_{1}\slashed{p}_1
	+C_{2}\slashed{p}_2\right)  + C_0 \slashed{p}_1  \slashed{\varepsilon}^* \slashed{q}\slashed{p}_2
	\crn&=C_{00} \left[ \varepsilon^*.q -(4-d)\slashed{\varepsilon}^* \slashed{q} \right]  + \left( C_{12} \slashed{p}_2+ C_{11} \slashed{p}_1 +  C_{1} \slashed{p}_1 \right)  \slashed{\varepsilon}^* \slashed{q}\slashed{p}_1 
	\crn&   + \left( C_{12} \slashed{p}_1+  C_{22} \slashed{p}_2 +  C_{1}\slashed{p}_1
	+C_{2}\slashed{p}_2  +C_{2}\slashed{p}_1 +C_{0}\slashed{p}_1 \right)   \slashed{\varepsilon}^* \slashed{q}\slashed{p}_2 .  
\end{align}
Because the divergent part  $C_{00}= \Delta_{\epsilon}/4 =1/(4\epsilon)$, which $d=4-2 \epsilon$, hence  $ C_{00}(4-d)=1/2$. The result is: 
\begin{align}
	\label{eq_sh2l1a}
	\slashed{k}_1\slashed{\varepsilon}^* \slashed{q} \slashed{k}_2  =& -\frac{1}{2}\slashed{\varepsilon}\slashed{q}  + \left[ C_{12} \left( \slashed{p}_1+ \slashed{q} \right)+ \left(  C_{11}  +  C_{1}  \right) \slashed{p}_1  \right]  \slashed{\varepsilon}^* \slashed{q} \left( \slashed{p}_2  -\slashed{q}\right) 
	\crn&   + \left[ \left( C_{12} +  X_0 \right) \slashed{p}_1 + \left( C_{22} 	+C_{2}   \right) \left( \slashed{p}_1+ \slashed{q} \right)  \right]    \slashed{\varepsilon}^* \slashed{q}\slashed{p}_2
	\crn=& -\frac{1}{2}\slashed{\varepsilon}^* \slashed{q}   +   X_{012}  \slashed{p}_1  \slashed{\varepsilon}^* \slashed{q}\slashed{p}_2, 
\end{align}
where we have used $ \varepsilon^*.q=q^2=0$ and  $\slashed{q} \slashed{\varepsilon}^* \slashed{q} =2 \varepsilon^*.q \slashed{q} - q^2 \slashed{\varepsilon}^*=0$.   The final result is 
\begin{align}
	\label{eq_sh2l1b}
	\overline{u_a}  m_F \left[A_1\right]  \left[ \slashed{k}_1\slashed{\varepsilon}^* \slashed{q} \slashed{k}_2 \right] u_{b} 
	%
	=& \overline{u_a} m_F   \left\{ \slashed{p}_1\slashed{\varepsilon}^* \left[ -\frac{1}{2} +m_b^2 X_{012}\right]  +\slashed{\varepsilon}^*\slashed{p}_2 \left[ -\frac{1}{2} +m_a^2 X_{012}\right]
	\right.\crn&\left.\qquad \quad 
	+(2p_1.\varepsilon^*) 
	 \left[ \frac{1}{2} - X_{012} \slashed{p}_1 \slashed{p}_2  \right]\right\} \left[A_1\right]     u_{b}. 
\end{align}
Consider the last two terms in the last line of the formula \eqref{eq_MA2a}
\begin{align}
	\label{eq_sh2l223}
	&		-k_2^2\slashed{k}_1\slashed{k} \slashed{\varepsilon}^*   - 	k_1^2\slashed{\varepsilon}^*\slashed{k} \slashed{k}_2
	\crn =& - k_2^2 \left( k^2  - \slashed{p}_1\slashed{k}\right) \slashed{\varepsilon}^* -\slashed{\varepsilon}^* \left( k^2 -\slashed{k} \slashed{p}_2 \right) k_1^2 
	\crn =& -k^2 \left( k_1^2+k^2_2\right) \slashed{\varepsilon}^* + \left( D_2 +m_V^2\right)\slashed{p}_1\slashed{k}\slashed{\varepsilon}^* +\left( D_1 +m_V^2\right) \slashed{\varepsilon}^* \slashed{k} \slashed{p}_2 
	\crn \to  &  - \slashed{\varepsilon}^*\frac{(D_0+ m_F^2)(D_1+D_2 +2m_V^2)}{D_0D_1D_2}+ \frac{\slashed{p}_1\slashed{k} \slashed{\varepsilon}^*}{D_0D_1}+  \frac{\slashed{\varepsilon}^* \slashed{k} \slashed{p}_2}{D_0D_2}  + m_V^2 \left(  \frac{\slashed{p}_1\slashed{k} \slashed{\varepsilon}^*}{D_0 D_1 D_2} + \frac{\slashed{\varepsilon}^* \slashed{k} \slashed{p}_2}{D_0D_1 D_2} \right)   
	\crn =& - \slashed{\varepsilon}^*\left[ 2m_V^2(B^{(0)}_0 +1) + 2m_V^2 B^{(0)}_0 +m_F^2 (B^{(1)}_0 +B^{(2)}_0 +2 m_V^2 C_0) \right]
	\crn&  -\slashed{p}_1 \left[ B^{(1)}_1\slashed{p}_1+  m_V^2 \left( C_1 \slashed{p}_1 + C_2 \slashed{p}_2\right) \right] \slashed{\varepsilon}^* - \slashed{\varepsilon}^* \left[ B^{(2)}_1\slashed{p}_2+  m_V^2 \left( C_1 \slashed{p}_1 + C_2 \slashed{p}_2\right) \right] \slashed{p}_2   
	\crn = & - \slashed{\varepsilon}^*\left[ m_V^2(4B^{(0)}_0 +2+2m_F^2  C_0+ m_a^2C_1 +m_b^2C_2) + m_a^2 B^{(1)}_1 +m_b^2 B^{(2)}_1 +m_F^2 (B^{(1)}_0 +B^{(2)}_0 )  \right]
	\crn& -m_V^2 \left(C_2 \slashed{p}_1\slashed{p}_2 \slashed{\varepsilon}^* + C_1 \slashed{\varepsilon}^* \slashed{p}_1\slashed{p}_2 \right) 
	\crn =&- \slashed{\varepsilon}^*\left[ m_V^2(4B^{(0)}_0 +2+2 m_F^2 C_0+ m_a^2C_1 +m_b^2C_2) + m_a^2 B^{(1)}_1 +m_b^2 B^{(2)}_1 +m_F^2 (B^{(1)}_0 +B^{(2)}_0 )  \right]
	\crn& +m_V^2X_3 \slashed{p}_1\slashed{\varepsilon}^*\slashed{p}_2+ (2p_1.\varepsilon^*) (-m_V^2) \left[ C_2 \slashed{p}_1 + C_1 \slashed{p}_2\right].
\end{align}

Lastly, consider the first term in the last line of the formula \eqref{eq_MA2a}: 
\begin{align}
	\label{eq_sh2l21}
	&2(k_1.\varepsilon^*) \slashed{k}_1\slashed{k} \slashed{k}_2 
	=  \left( k.\varepsilon^* - 2p_1.\varepsilon^* \right)  \times \left(\slashed{k} - \slashed{p}_1\right) \slashed{k} \left(\slashed{k} - \slashed{p}_2\right) 
	\crn= & \left(  - 2p_1.\varepsilon^* + 2 k.\varepsilon^*  \right)  \times \left( k^2\slashed{k} -k^2\slashed{p}_1 -k^2\slashed{p}_2 + \slashed{p}_1 \slashed{k}\slashed{p}_2\right) 
	\crn \to  &   \left(  - 2p_1.\varepsilon^* + 2 k.\varepsilon^*  \right)  \times \left[  \left( \frac{1}{D_1 D_2} + \frac{m_F^2}{D_0D_1 D_2} \right) \left( \slashed{k} -\slashed{p}_1 -\slashed{p}_2\right) + \frac{\slashed{p}_1 \slashed{k}\slashed{p}_2}{D_0 D_1 D_2} \right] 
	\crn \non= &   \left(  - 2p_1.\varepsilon^*  \right)  \times \left\{ \left[-\frac{1}{2}B^{(0)}_0 -m_F^2C_0  \right] (\slashed{p}_1 +\slashed{p}_2)  - m_F^2 (C_1\slashed{p}_1 +C_2\slashed{p}_2)  
	- \slashed{p}_1 ( C_1\slashed{p}_1 +C_2\slashed{p}_2)   \slashed{p}_2 \right\}	
	\crn&+   \left( 2 \varepsilon^*_{\mu}  \right)  \left[  \left( \frac{1}{D_1 D_2} + \frac{m_F^2}{D_0D_1 D_2} \right) \slashed{k}k^{\mu}  
	%
	-\left( \frac{1}{D_1 D_2} + \frac{m_F^2}{D_0D_1 D_2} \right) (\slashed{p}_1 +\slashed{p}_2)  k^{\mu} + \frac{\slashed{p}_1 \slashed{k}\slashed{p}_2 k^{\mu}}{D_0 D_1 D_2} \right]
	\crn=&  \left(  2p_1.\varepsilon^*  \right) \left\{ \left[ \frac{B^{(0)}_0}{2} + m_F^2C_0 \right] \left( \slashed{p}_1 +\slashed{p}_2\right) +  \left(m_F^2C_1 +m_{b}^2C_2 \right)\slashed{p}_1 +\left( m_F^2C_2 +m_{a}^2C_1\right)\slashed{p}_2 \right\}
	\crn&+   \left( 2 \varepsilon^*_{\mu}  \right)  \times \left\{  \left(B^{\mu\nu}+  m_F^2C^{\mu\nu} \right) \gamma_{\nu}   -\left(B^{\mu} + m_F^2C^{\mu} \right) \left( \slashed{p}_1 +\slashed{p}_2\right)  +C^{\mu \nu}\slashed{p}_1 \gamma_{\nu} \slashed{p}_2  \right\},
\end{align}
where  $B^{\mu} =B^{\mu}(0,m_V^2,m_V^2)$ and $B^{\mu\nu}=B^{\mu\nu}(0,m_V^2, m_V^2)$.  The last line in Eq. \eqref{eq_sh2l21} is expressed in terms of the PV functions as follows 
\begin{align}
	\label{eq_sh2l11a}
	&  \left( 2 \varepsilon^*_{\mu}  \right)  \left\{  \gamma_{\nu}  \left[\frac{g^{\mu\nu}}{2} \left( B^{(0)}_0 +1\right) +\frac{1}{6}B^{(0)}_0\left( 2p_1^\mu p_1^\nu +p_1^\mu p_2^\nu + p_2^\mu p_1^\nu +2 p_2^\mu p_2^\nu\right) \right]
	\right. \crn & \qquad \qquad\quad  +    m_F^2 \gamma_{\nu}   \left[ C_{00}g^{\mu\nu} +C_{11}p_1^\mu p_1^\nu +C_{12}p_1^\mu p_2^\nu + C_{12}p_2^\mu p_1^\nu +C_{22} p_2^\mu p_2^\nu \right] 
	\crn & \qquad \qquad\quad  - \left[ \frac{1}{2}B^{(0)}_0(p_1+p_2)^{\mu} - m_F^2 \left( C_1 p_1^{\mu} + C_2 p_2^{\mu}\right)\right] \left( \slashed{p}_1 +\slashed{p}_2\right) 
	\crn &\left.  \qquad \qquad\quad +  \left[ C_{00}g^{\mu\nu} +C_{11}p_1^\mu p_1^\nu +C_{12}p_1^\mu p_2^\nu + C_{12}p_2^\mu p_1^\nu +C_{22} p_2^\mu p_2^\nu \right]  \slashed{p}_1 \gamma_{\nu} \slashed{p}_2  \right\}
	\crn =& m_V^2\slashed{\varepsilon}^*  \left( B^{(0)}_0 +1\right) +(p_1.\varepsilon^*) B^{(0)}_0\left( \slashed{p}_1 +\slashed{p}_2\right) 
	\crn &   +    m_F^2   \left[ 2C_{00} \slashed{\varepsilon}^*  +\left( 2p_1. \varepsilon^* \right)    \left(C_{11} \slashed{p}_1 +C_{12}\slashed{p}_2 + C_{12}\slashed{p}_1 +C_{22} \slashed{p}_2 \right) \right] 
	\crn &   -  \left[  B^{(0)}_0 \left( 2 p_1.\varepsilon^*  \right) - m_F^2  \left( 2 p_1.\varepsilon^* \right) \left( C_1 + C_2 \right)\right] \left( \slashed{p}_1 +\slashed{p}_2\right) 
	\crn &+  \slashed{p}_1 \left[  2 \slashed{\varepsilon}^*  C_{00} +  \left( 2p_1.\varepsilon^* \right)  \left(C_{11}\slashed{p}_1 +C_{12}\slashed{p}_1+ C_{12}\slashed{p}_2+C_{22} \slashed{p}_2 \right) \right]    \slashed{p}_2.
\end{align}
Hence the final result of Eq. \eqref{eq_sh2l21} is 
\begin{align}
	\label{eq_38}
	2(k_1.\varepsilon^*) \slashed{k}_1\slashed{k} \slashed{k}_2 = &    \slashed{\varepsilon}^*  \left[ m_V^2(B^{(0)}_0 +1) + 2m_F^2 C_{00}  \right] + \slashed{p}_1   \slashed{\varepsilon}^* \slashed{p}_2 \left(2C_{00}\right) 
	\crn&+  (2p_1.\varepsilon^*)  \left[ m_F^2 X_{01}   +m_{b}^2X_2 \right] \slashed{p}_1 
	%
	+ (2p_1.\varepsilon^*)  \left[ m_F^2 X_{01}  +m_{a}^2 X_1 \right] \slashed{p}_2.
\end{align}
The sum of three terms given in Eqs. \eqref{eq_MA1b},  \eqref{eq_sh2l223}, and \eqref{eq_38}  gives $C_{(ab)L,R}$ corresponding to the diagrams (5)  given in Eqs. \eqref{eq_CLFVV} and \eqref{eq_CRFVV}.  The formulas of $D_{(ab)L,5}$ and $D_{(ab)R,5}$ are given in Eq. \eqref{eq_FVVDLhand}.

Regarding to the case of photon couplings in Eq. \eqref{eq_Gamma1}, the  equality given in Eq. \eqref{eq_Gak4} is still valid because the new part $ \Delta \Gamma_{\mu \alpha' \beta'} = \Gamma_{\mu \alpha' \beta'} -\Gamma'_{\mu \alpha' \beta'} =\delta k_v \left(g_{\mu \alpha'} q_{\beta'} -g_{\beta' \mu} q_{\alpha'} \right)$ satisfies  $\left(g_{\mu \alpha'} q_{\beta'} -g_{\beta' \mu} q_{\alpha'}\right) \varepsilon^{*\mu}  k_{1}^{\alpha'} k^{\beta'}_{2}= q^2(\varepsilon^*.k_2) - (q.k_2)(\varepsilon^*.q)=0$.   The other relevant part of $\mathcal{M}_5$ is:
\begin{align}
\label{eq_ADeGa}
&- \gamma_\alpha \left[ A \right]    
\gamma_\beta \times 
\Delta \Gamma_{\mu \alpha' \beta'} \varepsilon^{\mu*} \left( g^{\alpha \alpha'}g^{\beta \beta'}   -\frac{g^{\beta \beta'} k_1^\alpha k_1^{\alpha'} +g^{\alpha \alpha'} k_2^\beta k_2^{\beta'}}{m^2_V} \right)
\crn =&   \left( \slashed{q}\slashed{\varepsilon}^* -\slashed{\varepsilon}^*\slashed{q}\right) m_F  A_1 + \left( \slashed{q} \slashed{k} \slashed{\varepsilon}^* -\slashed{\varepsilon}^*\slashed{k}\slashed{q}\right)  A_2 
\crn&- \frac{1}{m_V^2} \left\{ \left[ (k_1.q) ( \slashed{k}_1 \slashed{\varepsilon}^* - \slashed{\varepsilon}^* \slashed{k}_2 ) + (k_1.\varepsilon^*) ( \slashed{q} \slashed{k}_2  - \slashed{k}_1 \slashed{q}  )\right] m_F A_1
\right. \crn &\left. \quad \quad \quad +\left[ (k_1.q) ( -\slashed{p}_1 \slashed{k} \slashed{\varepsilon}^* + \slashed{\varepsilon}^* \slashed{k} \slashed{p}_2 ) + (k_1.\varepsilon^*) ( \slashed{p}_1 \slashed{k}\slashed{q}  - \slashed{q} \slashed{k} \slashed{p}_2  ) \right]  A_2 \right\}.
\end{align}
 The final result of new contributions to  $i\delta \mathcal{M}_5$ is:
\begin{align}
	\label{eq_FVVa}
	i\delta \mathcal{M}_5 =&  \int \frac{d^4k}{(2\pi)^4} \frac{\overline{u_a}   \gamma_\alpha \left[ A \right]    
		\gamma_\beta u_{b} }{D_0 D_1 D_2} 
	%
	\times 
	\Delta \Gamma_{\mu \alpha' \beta'}  \left( g^{\alpha \alpha'}g^{\beta \beta'}   -\frac{g^{\beta \beta'} k_1^\alpha k_1^{\alpha'} +g^{\alpha \alpha'} k_2^\beta k_2^{\beta'}}{m^2_V} \right)
	\crn =& - \frac{i eQ_V \delta_v}{16\pi^2} \left\{ \frac{}{}\overline{u_a}   \left[  \left( 4p_1.\varepsilon^* - 2 \slashed{p}_1 \slashed{\varepsilon}^* -2\slashed{\varepsilon}^*\slashed{p}_2 \right) C_0 m_F A_1   \right] u_{b}
	\right.\crn&\left.  + \overline{u_a}   \left[ (2p_1.\varepsilon^*) (\slashed{p}_1 +\slashed{p}_2) -(m_a^2 +m_b^2) \slashed{\varepsilon}^* -2 \slashed{p}_1 \slashed{\varepsilon}^*\slashed{p}_2  \right] X_3A_2  u_{b}  
	\right.\crn&\left.  -\frac{1}{m_V^2} \overline{u_a}   \left[  C_{00}\left( \slashed{q}\slashed{\varepsilon}^* - \slashed{\varepsilon}^* \slashed{q}\right)  +(C_{11} +C_{12}) \left[-2(p_1.q) \slashed{\varepsilon}^* \slashed{p}_1  +2(p_1.\varepsilon^*) \slashed{q} \slashed{p}_1  \right]
	\right.\right. \crn&\left. \left. \qquad \qquad    +(C_{22} +C_{12}) \left[-2(p_2.q) \slashed{\varepsilon}^* \slashed{p}_2  +2(p_2.\varepsilon^*) \slashed{q} \slashed{p}_2  \right] 
	\right.\right. \crn&\left. \left. \qquad \qquad    + (p_1.q) \left[ -X_0 \left( -\slashed{p}_1\slashed{\varepsilon}^* + \slashed{\varepsilon}^* \slashed{p}_2 \right)  +(2p_1.\varepsilon^*) (C_2 -C_1) +2\left( C_1 \slashed{p}_1\slashed{\varepsilon}^* - C_2\slashed{\varepsilon}^* \slashed{p}_2 \right) \right] 
	\right.\right. \crn&\left. \left. \qquad \qquad   - (p_1. \varepsilon^*) \left( 2\slashed{p}_1\slashed{p}_2 -m_a^2 -m_b^2\right)(2X_3 +C_0) \right] m_F A_1  u_{b} 
\right.\crn&\left.  -\frac{1}{m_V^2} \overline{u_a}   \left[ \frac{}{}C_{00}\left(  2 (m_a^2 +m_b^2) \slashed{\varepsilon}^* + 4\slashed{p}_1\slashed{\varepsilon}^*\slashed{p}_2 - 2(\slashed{p}_1 +\slashed{p}_2 ) \times  (2 p_1.\varepsilon^*)\right) 
\right.\right. \crn&\left. \left. \qquad \qquad  + \frac{m_b^2 -m_a^2}{2}\times X_1 \left( -m_a^2  \slashed{\varepsilon}^*  - \slashed{p}_1\slashed{\varepsilon}^*\slashed{p}_2 +  (2 p_1.\varepsilon^*) \slashed{p}_1 \right)
\right.\right. \crn&\left. \left.  \qquad  \qquad  
+ \frac{m_b^2 -m_a^2}{2}\times X_2 \left(m_b^2  \slashed{\varepsilon}^*  + \slashed{p}_1\slashed{\varepsilon}^*\slashed{p}_2 -  (2 p_1.\varepsilon^*) \slashed{p}_2 \right) \right]A_2  u_{b} 
\right\}.
\end{align}
Ignoring the factor $ \frac{ eQ_V \delta k_v}{16\pi^2} $, the form factors are:
\begin{align}
\label{eq_dCLR}	
-\delta C^{FVV}_{(ab)L} =&  g^{LL} m_a \left[ X_3 + \frac{4C_{00} - (m_b^2 -m_a^2) X_1}{2m_V^2} \right] 
  +  g^{RR} m_b \left[ X_3 + \frac{4C_{00} +(m_b^2 -m_a^2) X_2}{2m_V^2} \right] 
  \crn& +  g^{RL} m_F \left[ 2 C_0 -\frac{8 C_{00} +m_a^2 (2X_1 +X_0) + m_b^2 (2X_2 +X_0)}{2 m_V^2} \right] 
  + g^{LR} \frac{m_F  m_am_bX_{012}}{m_V^2} ,
\crn - \delta C^{FVV}_{(ab)R} =&   \delta C_{L,5}\left[ g_{a}^{L} \leftrightarrow g_{a}^{R},  g_{b}^{L} \leftrightarrow g_{b}^{R} \right] ,
\crn -\delta D^{FVV}_{(ab)L} =&  g^{LL} \left[ -(m_a^2 +m_b^2)X_3 -\frac{4(m_a^2 +m_b^2)C_{00} +(m_b^2 -m_a^2) (-m_a^2 X_1 + m_b^2 X_2)}{2m_V^2}\right] 
\crn &+g^{RR} m_am_b \left[-2 X_3-\frac{8 C_{00} +(m_b^2 -m_a^2)(-X_1 +X_2)}{2m_V^2} \right] 
\crn &+g^{RL} m_a m_F\left[-2C_0 -\frac{-8 C_{00} +(m_b^2 -m_a^2)(2 X_1 +X_0)}{2m_V^2} \right] 
\crn& +g^{LR}m_bm_F\left[ -2 C_0  + \frac{8 C_{00} +(m_b^2 -m_a^2)(2 X_2 +X_0)}{2m_V^2} \right],
\crn \delta D^{FVV}_{(ab)R}= &    \delta D^{FVV}_{(ab)L} \left[ g_{a}^{L} \leftrightarrow g_{a}^{R},  g_{b}^{L} \leftrightarrow g_{b}^{R} \right]. 
\end{align}
All results given in Eq. \eqref{eq_dCLR} were cross checked using FORM package \cite{Vermaseren:2000nd}.  All formulas in Eq. \eqref{eq_dCLR}  satisfy automatically  the WI, namely   $\delta D^{FVV}_{(ab) L} +m_a \delta C^{FVV}_{(ab) L} + m_b\delta C^{FVV}_{(ab)R}=0$. 

\begin{center}
	\textbf{{Diagram (6)}}
\end{center}
After using the property of chiral operators $P_{L,R}$, the amplitude \eqref{eq_M6} is written as 
\begin{align}
	\label{eqM6a}
i\mathcal{M}_6 =& e Q_F \int \frac{d^4k}{(2\pi)^4}  
\frac{1}{D_0 D_1D_2} \times 
%
 \overline{u_{a}}  \left[   \left( m_F^2 \gamma_\alpha \slashed{\varepsilon}^{*}\gamma_\beta + \gamma_\alpha \slashed{k}_1\slashed{\varepsilon}^{*} \slashed{k}_2 \gamma_\beta \right) \left[ A_2\right] 
\right. \nonumber \\ & \left.  \qquad\; -  m_F \left[A_1\right] \left( g_{a}^{L^*}g_b^{R}P_R + g_{a}^{R^*}g_b^{L}P_L \right)  \left( \gamma_\alpha \slashed{k}_1\slashed{\varepsilon}^{*}  \gamma_\beta +  \gamma_\alpha \slashed{\varepsilon}^{*} \slashed{k}_2 \gamma_\beta \right) \right] u_{b}.	
\end{align} 
The numerator is divided into the two  parts  $N_1\sim g^{\alpha\beta}$ and $N_2\sim -k^{\alpha}k^{\beta}/m_V^2$. After extracting $g^{\alpha\beta}$, the first part is
\begin{align}
	\label{eq_TS1}
	N_1=&  \overline{u_a}  \left\{  \left[  (2-d) m_F^2  \slashed{\varepsilon}^{*} -2\slashed{k}_2\slashed{\varepsilon}^{*} \slashed{k}_1  + (4-d )  \slashed{k}_1\slashed{\varepsilon}^{*} \slashed{k}_2  \right]   \left[A_2\right] 
	\right. \nonumber \\
	& \left. \quad \;  -  m_F  \left[A_1\right]  \left[  4 \varepsilon^*.(k_1+k_2) -(4-d) \left(\slashed{k}_1\slashed{\varepsilon}^{*} +  \slashed{\varepsilon}^{*} \slashed{k}_2\right)   \right] \right\}  u_{b}. 
\end{align}
Ignoring the overall factor $eQ_F/(16\pi^2)$, the formula  in terms of tensor notations is  
\begin{align}
	\nonumber	N_1= & \overline{u_{a}}   \slashed{\varepsilon}^*   \left[A_2\right]  u_{b} \left[ -2 m_F^2 C_0 +(d-4)(d-2)C_{00}\right] + \left( 2 p_1.\varepsilon^*    \right)  \overline{u_a} \left[A_1\right] \left( 4m_F X_0   \right)   u_{b}
	\crn&+ \overline{u_a}  \left[ (2-d)  C_{\alpha \beta} \gamma^{\alpha}\slashed{\varepsilon}^{*} \gamma^{\beta}  + 2  C_{\alpha} \left(\slashed{p}_2\slashed{\varepsilon}^{*}\gamma^{\alpha} +  \gamma^{\alpha}\slashed{\varepsilon}^{*} \slashed{p}_1 \right)  -2 C_0 \slashed{p}_2\slashed{\varepsilon}^{*} \slashed{p}_1\right] \times	   \left[A_2\right]  u_{b} .
\end{align}
After expanding the tensors in terms of scalar PV-functions, the final result is
\begin{align} \label{eq_N1}
	N_1=& \overline{u_{a}}  \slashed{\varepsilon}^* \left[A_2\right]  u_{b} \left[ -2 m_F^2 C_0 +(d-2)^2C_{00} + 2 m_a^2 X_{01} +m_b^2 X_{02}\right]
	%
	+ \overline{u_{a}}     \slashed{p}_1 \slashed{\varepsilon}^* \slashed{p}_2\left[A_2\right]  u_{b} \times (2X_0)
	\crn&+\left( 2 \varepsilon^*. p_1  \right) \overline{u_a}  (-2) \left[ X_{01} \slashed{p}_1  
	%
	+ X_{02} \slashed{p}_2 \right]  \left[A_2\right] u_{b}
	%
	+ \left( 2 \varepsilon^*. p_1  \right)  \overline{u_a}  \left\{   4m_FX_0  \left[A_1\right] \right\} u_{b}.
\end{align}
Considering the second term proportional to $k^{\alpha}k^{\beta}$, we have
\begin{align}
	\label{eq_TS2}
	-m_V^2N_2 &= \overline{u_a}    \left( m_F^2  \slashed{k} \slashed{\varepsilon}^{*}\slashed{k}  + \slashed{k} \slashed{k}_1\slashed{\varepsilon}^{*} \slashed{k}_2\slashed{k} \right) \left[A_2\right]  u_{b}
	%
	-  m_F  \overline{u_a}  \left( \slashed{k} \slashed{k}_1\slashed{\varepsilon}^{*}\slashed{k} +  \slashed{k} \slashed{\varepsilon}^{*} \slashed{k}_2 \slashed{k}  \right) \left[A_1\right]  u_{b}. 
%
%
\end{align}
The two relations  $\slashed{k}\slashed{k}_1= D_1 +m_F^2 -m_{a}^2+\slashed{p}_1\slashed{k}$  and $\slashed{k}_2 \slashed{k} = D_2 +m_F^2 -m_{b}^2 +\slashed{k}\slashed{p}_2$ give 
\begin{align}
	\label{eq_TS21}
	N_2 \sim&  \overline{u_a}   \left( m_F^2  \slashed{k} \slashed{\varepsilon}^{*}\slashed{k}   \right)  \left[A_2\right]   u_{b}
	%
	+\overline{u_a}  \left(D_1 +m_F^2 -m_1^2  +\slashed{p}_1 \slashed{k}\right)\slashed{\varepsilon}^{*} \left(D_2 +m_F^2 -m_2^2  + \slashed{k}\slashed{p}_2 \right)  \left[A_2\right]  u_{b}
	\nonumber \\& -  m_F  \overline{u_a}  \left[   \left(D_1 +m_F^2 -m_a^2  +\slashed{p}_1 \slashed{k}\right) \slashed{\varepsilon}^{*}\slashed{k} +  \slashed{k} \slashed{\varepsilon}^{*}  \left(D_2 +m_F^2 -m_b^2  + \slashed{k}\slashed{p}_2 \right) \right]
	  \left[A_1\right]  u_{b}
	\crn \equiv & \overline{u_a} \left[   \left( L_1+L_2 \right) \left[A_2\right]   -m_F   \left[A_1\right]  L_3   \right]    u_{b},
\end{align}
where
\begin{align*}
	L_1 = &m_F^2 \left\{  \slashed{\varepsilon}^* \left[ (2-d)C_{00} - m_a^2 \left( C_{11} +C_{12}\right)  - m_b^2 \left( C_{12} +C_{22}\right) \right] 
	\right. \crn&\left.\quad\;  +(2p_1. \varepsilon^*) \left[ \left( C_{11} +C_{12}\right) \slashed{p}_1 + \left( C_{12} +C_{22}\right) \slashed{p}_2\right] \right\},
	\nonumber \\
	L_2 =& \frac{1}{D_0D_1D_2} \left(D_1 +m_F^2 -m_{a}^2  +\slashed{p}_1 \slashed{k}\right)\slashed{\varepsilon}^{*} \left(D_2 +m_F^2 -m_{b}^2  + \slashed{k}\slashed{p}_2 \right) 
	\nonumber\\ =& \slashed{\varepsilon}^{*} \left[  \frac{1}{D_0} + \frac{m_F^2 -m_{b}^2  + \slashed{k}\slashed{p}_2}{D_0D_2}  +  \frac{ m_F^2 -m_{a}^2 }{D_0D_1} +   \frac{ \left( m_F^2 -m_{a}^2\right) \left( m_F^2 -m_{b}^2\right)}{D_0 D_1 D_2} \right] +\frac{\slashed{p}_1 \slashed{k}\slashed{\varepsilon}^{*}}{D_0D_1} 
	\nonumber\\&+ 
	\frac{ \slashed{p}_1 \slashed{k} \slashed{\varepsilon}^{*} \slashed{k}\slashed{p}_2}{D_0D_1D_2} + \frac{  \slashed{p}_1 \slashed{k} \slashed{\varepsilon}^{*} \left( m_F^2 -m_{b}^2 \right)}{D_0D_1D_2}  + \frac{ \left( m_F^2 -m_{a}^2  \right) \slashed{\varepsilon}^{*} \slashed{k}\slashed{p}_2 }{D_0D_1D_2}
	\crn=& \slashed{\varepsilon}^{*}  \left[  A_0(m_V^2) + \left( m_F^2 -m_{b}^2\right) B^{(2)}_0 -m_b^2B^{(2)}_1   + \left( m_F^2 -m_{a}^2\right) B^{(1)}_0 -m_a^2B^{(1)}_1  
	\right.\crn& \left. +   \left( m_F^2 -m_{a}^2\right) \left( m_F^2 -m_{b}^2\right) C_0 \right]  
	\nonumber\\&+ 
	C_{\alpha\beta} (\slashed{p}_1 \gamma^{\alpha} \slashed{\varepsilon}^{*} \gamma^{\beta} \slashed{p}_2)  + C_{\alpha} (\slashed{p}_1 \gamma^{\alpha} \slashed{\varepsilon}^{*}) \left( m_F^2 -m_{b}^2 \right)    + C_{\alpha} ( \slashed{\varepsilon}^{*} \gamma^{\alpha}  \slashed{p}_2) \left( m_F^2 -m_{a}^2  \right),
	\crn L_3=& \frac{\slashed{\varepsilon}^{*} \slashed{k} }{D_0D_2} + \frac{\slashed{k} \slashed{\varepsilon}^{*}}{D_0D_1}+   \frac{m_F^2 (2k.\varepsilon^*) - m_{a}^2  \slashed{\varepsilon}^{*}\slashed{k}  -m_{b}^2\slashed{k}\slashed{\varepsilon}^{*} }{D_0D_1D_2} + \frac{  \slashed{p}_1 \slashed{k} \slashed{\varepsilon}^{*} \slashed{k} +\slashed{k} \slashed{\varepsilon}^{*} \slashed{k}\slashed{p}_2 }{D_0D_1D_2} 
	\crn =& -B^{(2)}_1\slashed{\varepsilon}^{*} \slashed{p}_2  -B^{(1)}_1\slashed{p}_1 \slashed{\varepsilon}^{*} - (2p_1.\varepsilon^*)(C_1 +C_2)m_F^2
	\crn& -C_{\alpha} \left( m_{a}^2 \slashed{\varepsilon}^{*}\gamma^{\alpha} + m_{b}^2\gamma^{\alpha}\slashed{\varepsilon}^{*} \right) +C_{\alpha \beta} \left( \slashed{p}_1 \gamma^{\alpha} \slashed{\varepsilon}^{*} \gamma^{\beta} + \gamma^{\alpha} \slashed{\varepsilon}^{*} \gamma^{\beta}\slashed{p}_2\right). 
\end{align*}
It can be proved that:
\begin{align}
\label{eq_interN2}	
	& C_{\alpha\beta} (\slashed{p}_1 \gamma^{\alpha} \slashed{\varepsilon}^{*} \gamma^{\beta} \slashed{p}_2) =   	\slashed{p}_1  \slashed{\varepsilon}^*  \slashed{p}_2 \left[ (2-d) C_{00} -m_a^2 (C_{11} +C_{12}) -m_b^2 (C_{22}+C_{12})\right]  
	\crn&\qquad \qquad \qquad  \qquad + (2p_1.\varepsilon^*) \left[ m_b^2 (C_{22}+C_{12})\slashed{p}_1 +m_a^2 (C_{11} +C_{12})\slashed{p}_2 \right] ,
\crn&  C_{\alpha} (\slashed{p}_1 \gamma^{\alpha} \slashed{\varepsilon}^{*})= -m_a^2 C_1\slashed{\varepsilon}^* -C_2 \left[ (2p_1.\varepsilon^*) \slashed{p}_1-	\slashed{p}_1  \slashed{\varepsilon}^*  \slashed{p}_2\right],
\crn  &C_{\alpha} ( \slashed{\varepsilon}^{*} \gamma^{\alpha} \slashed{p}_2) =   -m_b^2 C_2\slashed{\varepsilon}^* -C_1 \left[ (2p_1.\varepsilon^*) \slashed{p}_2-	\slashed{p}_1  \slashed{\varepsilon}^*  \slashed{p}_2\right],
\crn  & C_{\alpha} \left( m_{a}^2 \slashed{\varepsilon}^{*}\gamma^{\alpha} + m_{b}^2\gamma^{\alpha}\slashed{\varepsilon}^{*} \right)
\crn =&(2p_1.\varepsilon^*) \left[ -m_a^2 C_1 -m_b^2 C_2\right]  
%
+\slashed{p}_1  \slashed{\varepsilon}^*(m_a^2-m_b^2) C_1+   \slashed{\varepsilon}^*\slashed{p}_2 (m_b^2-m_a^2) C_2,
\crn& C_{\alpha \beta} \left( \slashed{p}_1 \gamma^{\alpha} \slashed{\varepsilon}^{*} \gamma^{\beta} + \gamma^{\alpha} \slashed{\varepsilon}^{*} \gamma^{\beta}\slashed{p}_2\right)
\crn =& (2p_1.\varepsilon^*) \left[ m_a^2 (C_{11} +C_{12}) +m_b^2 (C_{22} +C_{12}) +\slashed{p}_1\slashed{p}_2(C_{11} +2C_{12} +C_{22}) \right]  
\crn&+(\slashed{p}_1  \slashed{\varepsilon}^* +\slashed{\varepsilon}^*\slashed{p}_2) \left[ (2-d)C_{00} -m_a^2(C_{11} +C_{12}) -m_b^2(C_{22} +C_{12})\right]. \end{align}
Final results are:
\begin{align}
	L_1 = &m_F^2 \left\{\frac{}{}  \slashed{\varepsilon}^* \left[ (2-d)C_{00} - m_a^2 \left( C_{11} +C_{12}\right)  - m_b^2 \left( C_{12} +C_{22}\right) \right] 
\right. \crn&\left. +(2p_1. \varepsilon^*) \left[ \left( C_{11} +C_{12}\right) \slashed{p}_1 + \left( C_{12} +C_{22}\right) \slashed{p}_2\right] \right\},
%
	 \crn  L_2 =& \slashed{\varepsilon}^* \left\{ m_V^2(B^{(0)}_0 +1) + m_F^2(B^{(1)}_0 +B^{(2)}_0)  - m_a^2(B^{(1)}_0 +B^{(1)}_1) -  m_b^2(B^{(2)}_0 +B^{(2)}_1) 
	 \right.\crn&\left.\quad  +  m_F^4C_0 -m_F^2 \left[(m_a^2+m_b^2)C_0 +m_a^2C_1 +m_b^2C_2 \right] +m_a^2m_b^2 X_0\right\}
	 \crn& +\slashed{p}_1\slashed{\varepsilon}^*\slashed{p}_2 \left[ (2-d)C_{00} + m_F^2 X_3  -m_a^2X_1-m_b^2X_2\right]
	 \crn&+ (2p_1.\varepsilon^*) \left[  \left(  m_{b}^2 X_2  -m^2_FC_2\right) \slashed{p}_1 +\left( m_{a}^2X_1 -m^2_FC_1 \right)  \slashed{p}_2 \right],
	 %
	 %
		\nonumber\\L_3=& \slashed{p}_1\slashed{\varepsilon}^* \left[ -B^{(1)}_1 +(2-d) C_{00} -m_a^2 X_1 +m_b^2 (X_3 -X_2)\right]
\crn&+\slashed{\varepsilon}^*\slashed{p}_2 \left[ -B^{(2)}_1 +(2-d) C_{00} -m_b^2 X_2 +m_a^2 (X_3 -X_1)\right]
		\crn&+ (2p_1.\varepsilon^*) \left[ m_{a}^2 X_1  +m_{b}^2 X_2 -m_F^2X_3   +\slashed{p}_1\slashed{p}_2 \left( X_1 +X_2 -X_3\right) \right]. 
\end{align}
The above calculation is enough to derive  relevant contributions to  $C^{VFF}_{L,R}$ given in Eqs. \eqref{eq_VFFCL} and \eqref{eq_VFFCR}, and $D^{VFF}_{L,R}$ given in \eqref{eq_DLR6}. 

 \begin{center}
 	\textbf{Ward identity for the only gauge boson exchanges}
 \end{center}
Before coming to discuss the WI, we use the relations given in Eq. \eqref{eq_Xi} to write all the one-loop factors \eqref{eq_DabLR78}, \eqref{eq_FVVDLhand}, and \eqref{eq_DLR6}  from gauge boson exchanges in the following simple forms, ignoring the overall factor $e/(16\pi^2)$: 
\begin{align}
 D^{FV}_{(ab)L,78}
=& Q_e \left( g^{RL} m_a + g^{LR} m_b\right)(-3 m_FX_0) +Q_e  g^{RR} m_am_b \left[ -2 X^f_{12} + \frac{ \mathbf{b}^f_1 -m_F^2 (2X_0+X^f_{12}) }{m_V^2}  \right] 
\crn &+Q_e g^{LL} \left\{ \left( 2+ \frac{m_F^2+m_a^2 +m_b^2}{m_V^2} \right) \mathbf{b}^f_1 +1  + \frac{A_0(m_V^2) +m_a^2m_b^2 X^f_{12}}{m_V^2} 
\right. \crn&\qquad \qquad \quad\left. 
+ \frac{2 m_F^2(m_a^2 B^{(1)}_0 -m_b^2 B^{(2)}_0)}{(m_a^2 -m_b^2) m_V^2} \right\}. 
\end{align}
The WI for the $FVV$ and $FVV$ diagrams are  $f^{WI}_{FVV}\equiv D^{FVV}_{(ab)L}+ m_a C^{FVV}_{(ab)L} +m_b C^{FVV}_{(ab)R}$ and  $f^{WI}_{VFF}\equiv D^{VFF}_{(ab)L} + m_a C^{VFF}_{(ab)L} +m_b C^{VFF}_{(ab)R}$, respectively. 
The relations given in Eq. \eqref{eq_Xi} give:
\begin{align}
	\label{eq_rel2}	
	&2C^f_{00} +m_a^2X_1^f  +m_b^2X_2^f= -\mathbf{b}^f_1,
	\crn& X^v_{012}=-X^f_{12},
	\crn& m_F^2(X^v_{12} -X^v_3) +m_a^2 X_1^v +m_b^2 X_2^v +\mathbf{b}_1^v +1/2= m_F^2(X^v_{12} -X^v_3) -m_F^2C_0^v
	\crn&=m_F^2 (X^v_{012} -2X_0)=-m_F^2(X^f_{12}+2X_0),
	\crn &1/2 +m_a^2 X^v_1 +m_b^2 X^v_2 -m_F^2 X_3^v=(m_F^2 -m_V^2)X_0  -m_F^2C_0^v -m_F^2 X_3^v =-m_V^2X_0.
\end{align}
Combining the above formulas and results of $C_{i,ij}$ functions listed in Ref. \cite{Hue:2017lak}, the WI of all diagrams with boson exchanges is derived as follows 
\begin{align}
\label{eq_fIV}
f^{WI}_V = & D_{(ab)L,78} +f^{WI}_{FVV} +f^{WI}_{VFF}
\crn \sim&    \left( Q_e +Q_V -Q_F\right)
\crn  &\times \left\{ g^{LL} \left[  \frac{3 -B^{(0)}_0(m_V^2)}{2} +\frac{A_0(m_F^2)}{2m_V^2} 
 - \frac{(m_a^2 +m_F^2 -2m_V^2) m_V^2 +(m_a^2 -m_F^2)^2}{2(m_a^2 -m_b^2)m_V^2}  \times B^{(1)}_0 
\right. \right.\crn&\left. \left.\qquad\qquad  + \frac{(m_b^2 +m_F^2 -2m_V^2) m_V^2 +(m_b^2 -m_F^2)^2}{2(m_a^2 -m_b^2)m_V^2}  \times B^{(2)}_0  \right]
\right.\crn&\left. + g^{RR}  \left[   \frac{(m_V^2B^{(0)}_0 -A_0(m_F^2))\times (m_F^2 +2m_V^2) -(m_F^2 -4m_V^2)m_V^2}{2m_V^2}  
\right. \right.\crn&\left. \left.\qquad \quad - \frac{m_b^2 \left[ (m_a^2 +m_F^2 -2m_V^2) m_V^2 +(m_a^2 -m_F^2)^2\right]}{2(m_a^2 -m_b^2)m_V^2}   \times B^{(1)}_0  
\right. \right.\crn&\left. \left.\qquad\quad    +  \frac{ m_a^2 \left[ (m_b^2 +m_F^2 -2m_V^2) m_V^2 +(m_b^2 -m_F^2)^2\right] }{2 (m_a^2 -m_b^2)m_V^2} \times B^{(2)}_0  \right]
\right.\crn&\left. 
+\left( g^{RL} m_a +g^{LR} m_b\right) (3m_F X_0) \frac{}{}\right\}.
\end{align}
The final result is  $f^{WI}_V\sim Q_F -(Q_e+Q_V)=0$. In conclusion,  the contributions from the four diagrams with only gauge boson exchanges satisfy the WI when the electric charge conservation is valid.  
\section{ \label{app_WiFSV} Ward Identity for the diagrams of FSV-type  in the unitary gauge }
\begin{figure}[ht] 
	\centering
	\includegraphics[scale=0.8]{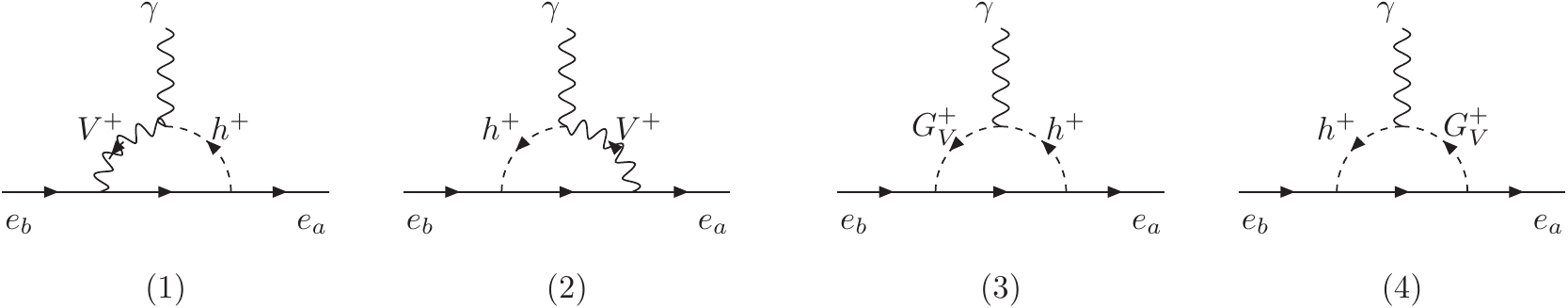}
	\caption{One-loop three-point FSV diagrams in the gauge $R_\xi$} \label{fig:FSV}
\end{figure}

This type of diagrams  were mentioned firstly in Ref. \cite{Yu:2021suw} for the general case of their contributions to  BSM. The  $\gamma-S-V$ vertices come the kinetic terms of the scalars:
\begin{align}
\label{eq_A-S-V}
L^D(S)= \left( \partial_{\mu}S -i P_{\mu}S\right)^{\dagger}\left( \partial^{\mu}S -i P^{\mu}S\right)= \left[g_{\gamma SV} g_{\mu \nu}S^{-Q} A^{\mu}V^{Q\nu}+ \mathrm{h.c.}\right] +\dots,
\end{align}
where $P_{\mu}$ containing the photon $A_{\mu}$ and $V_{\mu}$ is the covariant part of the covariant derivative of the Higgs multiplets. The Feynman diagrams in the general gauge $R_\xi$ are shown in Fig. \ref{fig:FSV}. 
Here only two diagrams (1) and (2) give non-zeros contributions in the unitary gauge, which correspond to the two diagrams (b) and (a) in Fig. 5 introduced in Ref. \cite{Yu:2021suw}. 
In this gauge,  the contributions of these two diagrams are: 
\begin{align}
\label{eq_9}
i\mathcal{M}_{9}
=& g_{\gamma SV}  \int \frac{d^4k}{(2\pi)^4} \times \frac{\overline{u_{a}}   [g_{a}^{L^*}P_R + g_{a}^{R^*}P_L] (m_F + \slashed{k}) \gamma_\alpha  [{g_{b}^{L}}P_L + {g_{b}^{R}}P_R] u_{b} }{D_0 D_1D_2}
%
\left( \varepsilon^{*\alpha} -\frac{(\varepsilon^*.k_{2}) k_{2}^{\alpha}}{m^2_V} \right) 
\crn
=& g_{\gamma SV}  \int \frac{d^4k}{(2\pi)^4} \times \frac{1}{D_0 D_1D_2}
\crn &\times \overline{u_{a}}\left\{ \slashed{\varepsilon}^* \left[ m_F \left[A_2\right] + \slashed{k} \left[A_1\right] \right] 
%
%
 -\frac{\slashed{k}_2 (k_2.\varepsilon^*)}{m_V^2}  m_F  \left[ A_2\right]
%
-\left[  A_1 \right]  
 \frac{\left( D_0 +m_F^2 -  \slashed{k} \slashed{p}_2 \right) (k_2.\varepsilon^*)}{m_V^2}   \right\}    u_{b}
\crn=& \frac{i g_{\gamma SV}}{16 \pi^2} \overline{u_{a}}\left\{ \slashed{\varepsilon}^* \left[ C_0 m_F \left[A_2\right] - (C_1 \slashed{p}_1 + C_2 \slashed{p}_2) \left[A_1\right] \right] 
\right.\crn&\left. -\frac{m_F}{m_V^2}  
\left[  (\gamma^{\mu} \varepsilon^{*\nu}) C_{\mu \nu} - (C_{\mu}\gamma^{\mu}) (p_2.\varepsilon^*) + \slashed{p}_2X_0 (p_1.\varepsilon^*)\right] \left[A_2\right]
\right.\crn&\left.   +\frac{1}{m_V^2} \left[A_1\right] \left[ C_{00}\slashed{\varepsilon}^*\slashed{p}_2 + (p_1.\varepsilon^*) \left( m_F^2 X_0 +X_1 \slashed{p}_1\slashed{p}_2 + m_b^2 X_2\right)\right] \right\} u_b,
\end{align}
where we have used $k_2.\varepsilon^*/(D_1D_2)\to 0$.  
The formulas of  $D_{L,R}$ and $C_{L,R}$ are:
\begin{align}
\label{eq_CD9}
eD^{Fhv}_{(ab)L,9}\times \left( \frac{g_{\gamma SV}}{16 \pi^2}\right)^{-1}=& g^{LL} m_F \left( C_0 - \frac{ C_{00}}{m_V^2}\right)   -g^{RL}  m_a C_1   +g^{LR} m_b \left( C_2 + \frac{C_{00}}{m_V^2}\right),
\crn D^{Fhv}_{(ab)R,9}
=& D^{Fhv}_{(ab)L,9} \left[g^{L}_a \leftrightarrow g^{R}_a, g^{L}_b \leftrightarrow g^{R}_b \right], 
\crn eC^{Fvh}_{(ab)L,9}\times \left( \frac{g_{\gamma SV}}{16 \pi^2}\right)^{-1}= &  -g^{RL}  C_2 - \frac{m_F}{2m_V^2} \left[ g^{LL} m_a X_1 + g^{RR} m_b X_{02}  \right]
\crn& +\frac{1}{2m_V^2} \left[ g^{RL} (m_F^2 X_0 +m_b^2 X_2) + g^{LR}  m_a m_b X_1  \right],
\crn C^{Fhv}_{(ab)R,9}= &C_{(ab)L} \left[g^{L}_a \leftrightarrow g^{R}_a, g^{L}_b \leftrightarrow g^{R}_b \right], 
\end{align}
where $X_i^{Fhv}\equiv X_i(m_a^2,0,m_b^2; m_F^2,m_h^2,m_V^2)$. 
Similarly, the results for  diagram (10) are: 
\begin{align}
\label{eq_CD10}
eD^{Fvh}_{(ab)L,10}\times \left( \frac{g_{\gamma SV}}{16 \pi^2}\right)^{-1}=& g^{LL} m_F \left( C_0 - \frac{ C_{00}}{m_V^2}\right)   -g^{LR}  m_b C_2   +g^{RL}  m_a \left( C_1 + \frac{C_{00}}{m_V^2}\right),
\crn D^{Fvh}_{(ab)R,10}
 =&  D^{Fvh}_{(ab)L,10} \left[g^{L}_a \leftrightarrow g^{R}_a, g^{L}_b \leftrightarrow g^{R}_b \right], 
\crn eC^{Fvh}_{(ab)L,10}\times \left( \frac{g_{\gamma SV}}{16 \pi^2}\right)^{-1}= & - g^{R*}_a g^{L}_b  C_1  - \frac{m_F}{2m_V^2} \left[ g^{R*}_a g^{R}_b m_b X_2 + g^{L*}_a g^{L}_b m_a X_{01}  \right]
\crn& +\frac{1}{2m_V^2} \left[ g^{R*}_a g^{L}_b (m_F^2 X_0 +m_a^2 X_1) + g^{L*}_a g^{R}_b m_a m_b X_2  \right],
\crn eC^{Fvh}_{(ab)R,10}\times \left( \frac{g_{\gamma SV}}{16 \pi^2}\right)^{-1}= & C^{Fvh}_{(ab)L} \left[g^{L}_a \leftrightarrow g^{R}_a, g^{L}_b \leftrightarrow g^{R}_b \right], 
\end{align}
where $X^{Fvh}_{i} \equiv X_i(m_a^2,0,m_b^2; m_F^2,m_V^2,m_h^2)$.  The above formulas are  consistent with calculation using  FORM. 
 The corresponding formulas of WI are 
\begin{align}
	\label{eq_WIfhv}	
	\frac{f_{WI}^{Fhv}}{k_{\gamma SV}} &=  g^{LL}  m_F \left[  2(m_V^2 C_0 -C_{00}) +m_a m_b X_{012} \right]^{fhv} -  g^{RR}  m_F  \left[ m_a^2 X_{1} +m_b^2 X_{02} \right]^{fhv} 
	\crn &+ g^{RL}  \left[ -2 m_V^2 (m_a C_1 +m_b C_2)  +m_b \left( m_F^2 X_0 +m_a^2 X_1 +m_b^2 X_2\right) \right]^{fhv} 
	\crn &
	+ g^{LR} \left[ 2 m_V^2 (m_b C_2 -m_a C_1)+ 2 m_b C_{00}   +m_a \left( m_f^2 X_0 +m_b^2 X_{12}\right) \right]^{fhv},
\crn  \frac{f_{WI}^{Fvh}}{k_{\gamma SV}} &=   g^{LL} m_F\left[ 2(m_V^2 C_0 -C_{00}) -m_a m_b X_{012} \right] -  g^{RR} m_F \left[ m_a^2 X_{01} +m_b^2 X_{2} \right]^{fvh} 
	\crn &
	+ g^{RL} \left[ 2 m_V^2 (m_a C_1 -m_b  C_2)  + 2 m_a C_{00} +m_b \left( m_f^2 X_0 + m_a^2 X_{12} \right) \right]^{fvh}
	\crn &+ g^{LR} \left[ -2 m_V^2 (m_a C_1 +m_b C_2)  +m_a \left( m_f^2 X_0 +m_a^2 X_1 +m_b^2 X_2\right) \right]^{fvh},  
\end{align}
where $k_{\gamma SV}=g_{\gamma SV}/(32 \pi^2 m_V^2)$. The WI is valid  if only  $f_{WI}^{Fhv}+ f_{WI}^{Fvh}=0$. We can see crudely that all $C_{(ab)L,9}$, $C_{(ab)R,9}$, $C_{(ab)L,10}$, and  $C_{(ab)R,10}$ are convergent. In contrast, all $D_{(ab)L,9}$, $D_{(ab)R,9}$, $D_{(ab)L,10}$, and  $D_{(ab)R,10}$ contain divergent terms. Therefore, the necessary condition  to guarantee the validation of the WI given in Eq. \eqref{eq_DLR} is that all of these divergent terms must vanish. Strictly,   the WI is valid  if only  $g_{\gamma SV}$=0 or $g^L_a=g^R_a=0$. Because  at least one of $g^L_a$ or $g^R_a$ must be non-zero, the condition $g_{\gamma SV}$=0 is the only valid choice, i.e.,   the vertex-type $\gamma$-$S$-$V$ does not appear in the all BSM guaranteeing the WI for the external photon. This conclusion is also true for the case $a=b$, corresponding to  the one-loop contribution to the AMM of the leptons. 

Finally, using the assumption of the Lagrangian for couplings of the Goldstone boson given in Eq. \eqref{eq_Gvcoupling}, we can determine the one-loop contributions of the FSV diagrams mentioned above, using the general gauge $R_\xi$. The propagator of the  gauge boson $V$  can be written in terms of two separated parts:
\begin{align}
	\Delta^{(\xi) \mu\nu}_{V}(k^2)&\equiv \Delta^{(u)\mu\nu}_{V}(k^2)+ \Delta^{(T) \mu\nu}_{\xi, V }(k^2),
	\crn \Delta^{(u) \mu\nu}_{V}(k^2) &=\frac{-i}{k^2-m^2_V}\left(g^{\mu\nu}-\frac{k^{\mu} k^{\nu}}{m^2_V}\right),
	\crn  \Delta^{(T) \mu\nu}_{\xi, V}(k^2) &=\frac{-i}{ m^2_V}\times  \frac{k^{\mu} k^{\nu}}{k^2-\xi\, m^2_V}=  \frac{-k^\mu k^{\nu}}{m_V^2} \times i \Delta^{0\xi}_{G_V},  \label{pxi_gaugeV}
\end{align} 
where $ \Delta^{(u) \mu\nu}_{V}(k^2)$  is the propagator in the unitary gauge, and  $\Delta^{0\xi}_{G_V}$ relates to the propagator of $G_V$ as follows:
\begin{align}
	\label{eq_DeltaGV}
	\Delta^{\xi}_{G_V}& = i \Delta^{0\xi}_{G_V}=  \frac{i}{k^2- \xi m_V^2}	=\left[\begin{array}{cc}
		0&  \xi \to \infty: \;\mathrm{Unitary} (u), \\
		\frac{i }{k^2-m_V^2},	&  \xi =1: \;\mathrm{'t\; Hooft-Feynman} (HF)
	\end{array}\right.. 	
\end{align}
For two diagrams (3) and (4) in Fig. \ref{fig:FSV}, the Feynman rules for the couplings $\gamma -S-G_V$ are the same form as those given in Lagrangian \eqref{eq_AX12}, namely $S\equiv h_1$ and $G_V\equiv h_2$. The reason is that all mass eigenstates of the scalar with the same electric charges  come from the same squared mass matrix.   Therefore,  $\mathcal{L}^{\gamma hG_V}= ieQ_HA^{\mu} \left[ \left( h^* \partial_{\mu}G_V -G_V \partial_{\mu}h^*\right) +\mathrm{h.c.}\right] $. Formulas corresponding to diagrams (1) and (3) of Fig. \ref{fig:FSV} in the general gauge $R_\xi$ are 
\begin{align*}
	i\mathcal{M}^{(\xi)}_{9}
	=&	i\mathcal{M}^{(u)}_{9}
	\crn&+  g_{\gamma SV}  \int \frac{d^4k}{(2\pi)^4} \times \frac{\overline{u_{a}}   [g_{a,Fh}^{L^*}P_R + g_{a,Fh}^{R^*}P_L] (m_F + \slashed{k}) \gamma_\alpha  [{g_{b,FV}^{L}}P_L + {g_{b,FV}^{R}}P_R] u_{b} (\varepsilon^*.k_{2}) k_{2}^{\alpha} }{D_0 D_1D_2 m^2_V}, 
	%
\end{align*}
where $D_0=k^2-  m_{F}^2$,  $D_1=k_1^2-  m_{h}^2$, $D_2=k_2^2-\xi m^2_{V}$, and $\mathcal{M}^{(u)}_{9}$ is exactly the part  given in Eq. \eqref{eq_9}, calculated in the unitary gauge. 
The results of the two diagrams (1) and (3) are:
\begin{align}
	i\Delta \mathcal{M}^{(\xi)}_{9}
	\equiv &i\mathcal{M}^{(\xi)}_{9}-	i\mathcal{M}^{(u)}_{9}
	\crn=&  \frac{i g_{\gamma SV}}{16 \pi^2} \overline{u_{a}}\left\{  \frac{m_F}{m_V^2}  
	\left[  (\gamma^{\mu} \varepsilon^{*\nu}) C_{\mu \nu} - (C_{\mu}\gamma^{\mu}) (p_2.\varepsilon^*) + \slashed{p}_2X_0 (p_1.\varepsilon^*)\right] \left[A_2\right]
	\right.\crn&\left.   +\frac{1}{m_V^2} \left[A_1\right] \left[ C_{00}\slashed{\varepsilon}^*\slashed{p}_2 + (p_1.\varepsilon^*) \left( m_F^2 X_0 +X_1 \slashed{p}_1\slashed{p}_2 + m_b^2 X_2\right)\right] \right\} u_b,
	\label{eq_M1a}
	\\ i\mathcal{M}^{(\xi)}_3  
	= &  \frac{-ieQ_H}{16\pi^2 } \overline{u_{a}}\left\{ -2p_1.\varepsilon^*  \left[A_1 \right]    m_F  X_0  
	+  \left[ 2C^f_{00}\slashed{\varepsilon}^* + \left( X_1^f \slashed{p}_1 + X_2^f \slashed{p}_2 \right) (2p_1.\varepsilon^*) \right]
	[A_2]    \right\}  u_{b}, \label{eq_M3hG}
\end{align}
where $C_{00}=C_{00}(m_F^2, m_h^2,\xi m_V^2)$ and $X_{0,i}= X_{0,i}(m_F^2, m_h^2,\xi m_V^2)$. 

As we showed clearly in Eq. \eqref{pxi_gaugeV}, a  propagator  of an arbitrary internal gauge boson always consists of   two parts: i) the first part is exactly the unitary propagator resulting in $\mathcal{M}^{(u)}_{9}$, and ii) the second is proportional to the propagator of the respective Goldstone boson, in which the  parameter $\xi$ defines a new mass value in the denominator, which results in $\Delta \mathcal{M}^{(\xi)}_{9} +\mathcal{M}^{(\xi)}_3$. Because $\xi$ is arbitrary, the two one-loop contributions corresponding to the two mentioned parts are independent.  As a result, the WI violation of the contributions relating to $\mathcal{M}^{(u)}_{9}$ is enough to guarantee that the contributions from the FSV-type diagrams always violate the WI. 

\section{Higgs gauge couplings in the Higgs triplet models}
Here we summarize the HTM  and derive precisely the Higgs gauge couplings.  The Higgs sector consists of a Higgs triplet $\Delta \sim (3,2)$ and a Higgs doublet $\Phi\sim (2,1)$ in the electroweak gauge symmetry $SU(2)_L\times U(1)_Y$ corresponding to the electric operator $Q=T^3+Y/2$. Here we will use the notations from Ref. \cite{Ashanujjaman:2021txz, Aoki:2011pz}, the Higgs sector is 
\begin{align}
	\label{eq_Higgs}
	\Phi= \begin{pmatrix}
		\varphi^+	\\
		\frac{1}{\sqrt{2}} \left(\varphi +v_{\Phi} +i\chi\right) 
	\end{pmatrix}, \; \Delta =\begin{pmatrix}
		\frac{\Delta^+}{\sqrt{2}}&  \Delta^{++}\\
		\Delta^0& -\frac{\Delta^+}{\sqrt{2}} 
	\end{pmatrix} \; \mathrm{with }\; \Delta^0= \frac{\delta +v_{\Delta} +i \eta}{\sqrt{2}},		
\end{align}
where $v_{\Phi}$ and $v_{\Delta}$ are the vacuum expectation values (VEV) of the neutral Higgs components. Because $v_{d}$ has the lepton number 2, $v_d\ll v_{\Delta}$. 

The Higgs gauge couplings appear in the following kinetic terms:
\begin{align}
	\label{eq_LkH}
	\mathcal{L}_{k,H}&= \left(D_{\mu} \Phi\right)^{\dagger} \left(D^{\mu} \Phi\right) + \mathrm{Tr}\left[\left(D_{\mu} \Delta\right)^{\dagger} \left(D^{\mu} \Delta\right)\right], 
\end{align}
where
\begin{align}
	\label{eq_Dmu}
	D_{\mu} \Phi= \left(\partial_{\mu} + i\frac{g}{2} \tau^a W^a_{\mu} +i \frac{g'}{2}B_{\mu}\right) \Phi, \;  D_{\mu} \Delta= \partial_{\mu} \Delta + i\frac{g}{2}\left[\tau^a W^a_{\mu},\Delta\right]  +i \frac{g'}{2}B_{\mu} \Delta. 
\end{align}
The masses and mixing parameters  of the gauge bosons are  derived from the Eq. \eqref{eq_LkH}, with VEVs of $\Phi$ and $\Delta$. A detailed calculation shows that the physical states  $W^\pm$, neutral $Z$ and photon $A_{\mu}$ are:
\begin{align}
	\label{eq_gaug}
	W^\pm_{\mu}=\frac{W^1_{\mu} \mp iW^2_{\mu}}{\sqrt{2}}, \; W^3_{\mu}=c_W Z_{\mu} +s_W A_{\mu},\; 	B_{\mu}= -s_W Z_{\mu} +c_W A_{\mu}.
\end{align}
The respective masses are $m_W^2=g^2(v^2_{\Phi} +2v^2_{\Delta})/4$, $m_Z^2=g^2(v^2_{\Phi} +2 v^2_{\Delta})/(4 c^2_W)$ and the photon is massless. The relation $g'/g=t_W$ is well-known, the same as that  in the SM. 
The covariant derivatives in Eq. \eqref{eq_Dmu} are written in the mass eigenstates as follows:
\begin{align}
	\label{eq_VVH}
	D_{\mu}\Delta &=		\frac{ig}{2}\begin{pmatrix}
		v_{\Delta} W^+_{\mu} +\sqrt{2}t_W B_{\mu},& -2 \Delta^+ W^+_{\mu} 	\\
		2 \Delta^+W^-_{\mu} - \frac{\sqrt{2} v_{\Delta}Z_{\mu}}{c_W},& -v_{\Delta} W^+_{\mu} -\sqrt{2} t_W \Delta^+ B_{\mu} 
	\end{pmatrix} +\dots ,
	\crn D_{\mu}\Phi&= \frac{ig}{2}	\begin{pmatrix}
		\left( \frac{c^2_W -s^2_W}{c_W} Z_{\mu} +2 s_W A_{\mu}\right) \varphi^+	+ v_{\Phi} W^+_{\mu}\\
		\sqrt{2}W^-_{\mu} \varphi^+ -\frac{v_{\Phi}}{\sqrt{2}c_W}Z_\mu
	\end{pmatrix}+ \dots,
\end{align}
where we just focus on the couplings $SVV$ relating to the vertex $H^\pm W^\mp \gamma$. Therefore, the relevant parts in the kinetic term are:
\begin{align}
	\mathcal{L}_{k,H} &= \frac{g^2}{4} \left( 2 s_W A_{\mu} \varphi^-	+ v_{\Phi} W^-_{\mu}\ \right)	\left( 2 s_W A^{\mu} \varphi^+	+ v_{\Phi} W^{+\mu}\ \right)
	\crn&	+  \frac{g^2}{2} \left(v_{\Delta} W^-_{\mu} +\sqrt{2}t_W\Delta ^- B_{\mu}\right) \left(v_{\Delta} W^{+\mu} +\sqrt{2}t_W\Delta ^+ B^{\mu}\right) +\dots 
	\crn&= \frac{g^2s_W}{2} \left[ \left(v_{\Phi} \varphi^- + \sqrt{2}  v_{\Delta} \Delta^-\right) W^{+\mu} +\mathrm{h.c.}\right] A_{\mu} + \dots
\end{align}
The Higgs potential of all Higgs multiplets was investigated previously, for example, \cite{Ashanujjaman:2021txz, Aoki:2011pz}. The results of masses and mixing parameters of all Higgs bosons are confirmed by our careful cross-check. We focus on the Higgs gauge couplings of the singly charged Higgs boson in this model,  the mixing parameter $\beta_\pm$ relating to mass eigenstates and the original ones are:
\begin{align}
	\label{eq_betapm}
	\begin{pmatrix}
		\varphi^\pm	\\
		\Delta^\pm	
	\end{pmatrix} =\begin{pmatrix}
		c_{\beta_\pm}&  -s_{\beta_\pm}\\
		s_{\beta_\pm}& c_{\beta_\pm}
	\end{pmatrix} 	\begin{pmatrix}
		G^\pm_W	\\
		H^\pm	
	\end{pmatrix}, \; t_{\beta_\pm}= \frac{\sqrt{2}v_\Delta}{v_{\Phi}}. 
\end{align}
Here $G^\pm_W$ is the Goldstone bosons of $W^\pm$, while $H^\pm$ is the only singly charged Higgs boson predicted by the HTM. 
Then the couplings $H^\pm W^\mp\gamma\sim \sqrt{2}c_{\beta_\pm} v_{\Delta} -s_{\beta_\pm} v_{\Phi}=0$, and the couplings with $G^\pm_W$ are $(e m_W)\left[  W^{+}_{\mu} G^-_W +\mathrm{h.c.}\right] A_{\mu}$, consistent with the SM. In contrast,  Ref. \cite{Yu:2021suw}  seems to take into account only the contribution of $\Delta^\pm$ to $H^\pm$, and ignored that of  $\varphi^\pm$, although they have the same amplitude but opposite signs.

It is noted that the results derived from our calculation are consistent with those in recent works discussing all tree-level  decays of Higgs and gauge bosons predicted by the HTM at LHC \cite{Ashanujjaman:2021txz, Aoki:2011pz}. The decays $H^\pm\to W^\pm  \gamma$ do not appear in the decay lists of these works.

\end{document}